\documentclass[onecolumn,10pt]{IEEEtran}
\usepackage{ifpdf, flushend}
\ifCLASSINFOpdf
  \usepackage[pdftex]{graphicx}
  \graphicspath{{../pdf/}{../jpeg/}}
  \DeclareGraphicsExtensions{.pdf,.jpeg,.png}
\else
  \usepackage[dvips]{graphicx}
  \graphicspath{{../eps/}}
  \DeclareGraphicsExtensions{.eps}
\fi
\usepackage[cmex10]{amsmath}
\usepackage {amssymb}
\usepackage{algorithmic}
\usepackage{array}
\usepackage{mdwmath}
\usepackage{mdwtab}
\usepackage{eqparbox}
\usepackage{hyperref}
\usepackage{algorithm}
\usepackage{algorithmic}

\newcommand{\argmax}{\operatornamewithlimits{argmax}}
\newcommand{\beq}{\begin{equation}}
\newcommand{\eeq}{\end{equation}}
\newcommand{\beqn}{\begin{eqnarray}}
\newcommand{\eeqn}{\end{eqnarray}}
\newcommand{\beqno}{\begin{eqnarray*}}
\newcommand{\eeqno}{\end{eqnarray*}}
\newcommand{\bma}{\begin{displaymath}}
\newcommand{\ema}{\end{displaymath}}
\newcommand{\bnu}{\begin{enumerate}}
\newcommand{\enu}{\end{enumerate}}
\newcommand{\bce}{\begin{center}}
\newcommand{\ece}{\end{center}}
\newcommand{\btb}{\begin{tabular}}
\newcommand{\etb}{\end{tabular}}

\begin{document}
%
% paper title
% can use linebreaks \\ within to get better formatting as desired
\title{Distributed MAC Protocol for Cognitive Radio Networks: Design, Analysis, and Optimization}

\author{\IEEEauthorblockN{Le Thanh Tan and Long Bao Le}  %\vspace{-0.2cm}
\thanks{The authors are with INRS-EMT, University of Quebec,  Montr\'{e}al, Qu\'{e}bec, Canada. 
Emails: \{lethanh,long.le\}@emt.inrs.ca. \textbf{L. B. Le} is the corresponding author.}}

% make the title area
\maketitle

\begin{abstract}
%\boldmath
In this paper, we investigate the joint optimal sensing and distributed MAC protocol design problem for cognitive radio networks.
We consider both scenarios with single and multiple channels.
For each scenario, we design a synchronized MAC protocol for dynamic spectrum sharing among multiple secondary users, which
incorporates spectrum sensing for protecting active primary users. 
We perform saturation throughput analysis for the corresponding proposed MAC protocols that 
explicitly capture spectrum sensing performance. Then, we find their optimal
configuration by formulating throughput maximization problems subject to detection probability constraints 
for primary users. In particular, the optimal solution of the optimization problem returns the required sensing time for
primary users' protection and optimal contention window for maximizing total throughput of the secondary network.
Finally, numerical results are presented to illustrate developed theoretical findings in the paper and significant performance
gains of the optimal sensing and protocol configuration.
\end{abstract}

\begin{IEEEkeywords}
MAC protocol, spectrum sensing, optimal sensing, throughput maximization, cognitive radio.
\end{IEEEkeywords}
%\IEEEpeerreviewmaketitle

\section{Introduction}
Emerging broadband wireless applications have been demanding unprecedented increase in radio spectrum resources.
As a result, we have been facing a serious spectrum shortage problem. However, several recent
measurements reveal very low spectrum utilization in most useful frequency bands \cite{Zhao07}.
To resolve this spectrum shortage problem, the Federal Communications Commission (FCC) has opened licensed bands for 
unlicensed users' access. This important change in spectrum regulation has resulted in growing research interests on dynamic 
spectrum sharing and cognitive radio in both industry and academia. In particular, IEEE has established an IEEE 802.22 workgroup to 
build the standard for WRAN based on CR techniques \cite{R9}.

Hierarchical spectrum sharing between primary networks and secondary networks is one of the most widely studied dynamic
spectrum sharing paradigms. For this spectrum sharing paradigm, primary users typically have strictly higher priority 
than secondary users in accessing the underlying spectrum. One potential approach for dynamic spectrum sharing
is to allow both primary and secondary networks to transmit simultaneously on the same frequency with appropriate interference control
to protect the primary network \cite{Kim081}, \cite{Le08}. In particular, it is typically required that a certain
interference temperature limit due to secondary users' transmissions must be maintained at each primary receiver.
Therefore, power allocation for secondary users should be carefully performed to meet stringent interference requirements in this 
spectrum sharing model.

Instead of imposing interference constraints for primary users,  spectrum sensing can be adopted by secondary users to search for
and exploit  spectrum holes (i.e., available frequency bands) \cite{R1, Peh09}. There are several challenging technical issues 
related to this spectrum discovery and exploitation problem. On one hand, secondary users should spend sufficient time 
for spectrum sensing so that they do not interfere with active primary users. On the other hand, secondary users should efficiently
exploit spectrum holes to transmit their data by using an appropriate spectrum sharing mechanism. Even though these aspects are tightly
coupled with each other, they have not been treated thoroughly in the existing literature.

In this paper, we make a further bold step in designing, analyzing, and optimizing MAC protocols for cognitive radio networks
considering sensing performance captured in detection and false alarm probabilities. 
Specifically, the contributions of this paper can be summarized as follows: \emph{i}) we design distributed synchronized MAC protocols for
cognitive radio networks incorporating spectrum sensing operation for both single and multiple channel scenarios; \emph{ii}) we analyze 
saturation throughput of the proposed MAC protocols; 
\emph{iii}) we perform throughput maximization of the proposed MAC protocols against their key parameters, namely sensing time and minimum
contention  window; \emph{iv}) we present numerical results to illustrate performance of the proposed
MAC protocols and the throughput gains due to optimal protocol configuration.
 
The remaining of this paper is organized as follows. Section ~\ref{SystemModel} describes system and sensing models. MAC protocol design,
throughput analysis, and optimization for the single channel case are performed in Section ~\ref{SingleChan}.
The multiple channel case is considered in Section~\ref{MultipleChan}. Section ~\ref{Results} presents 
numerical results followed by concluding remarks in Section ~\ref{conclusion}.

\section{Related Works}

Various research problems and solution approaches have been considered for a dynamic spectrum sharing problem in
 the literature.  In \cite{Kim081}, \cite{Le08}, a dynamic power allocation problem for cognitive radio
networks was investigated considering fairness among secondary users and interference constraints for primary users. 
When only mean channel gains averaged over short term fading can be estimated, the authors proposed more relaxed
protection constraints in terms of interference violation probabilities for the underlying fair power allocation problem.
In \cite{Dev06}, information theory limits of cognitive radio channels were derived. Game theoretic approach for 
dynamic spectrum sharing was considered in \cite{wang08}, \cite{niyato08}. 

There is a rich literature on spectrum sensing for cognitive radio networks (e.g., see \cite{Yu09} and references therein).
Classical sensing schemes based on, for example, energy detection techniques or advanced cooperative sensing strategies \cite{Un08} where multiple
secondary users collaborate with one another to improve the sensing performance have been investigated in the literature.
 There are a large number of papers
considering MAC protocol design and analysis for cognitive radio networks \cite{Cor09}-\cite{Su08} (see \cite{Cor09} for a survey 
of recent works in this topic). However, these existing works either assumed perfect spectrum sensing or did not explicitly model the sensing imperfection in their design and analysis. In \cite{R1}, optimization of sensing and throughput tradeoff under a detection probability constraint was investigated. It was shown that the detection constraint is met with equality at optimality. However, this optimization tradeoff was only investigated
for a simple scenario with one pair of secondary users. Extension of this sensing and throughput tradoff to wireless fading channels 
was considered in \cite{zuo10}.

\textbf{
There are also some recent works that propose to exploit cooperative relays to improve sensing and throughput performance
of cognitive radio networks. In particular, a novel selective fusion spectrum sensing and best relay data transmission
scheme was proposed in \cite{zuo111}. Closed-form expression for the spectrum hole utilization efficiency of the
proposed scheme was derived and significant performance improvement compared to other sensing and transmission schemes
was demonstrated through extensive numerical studies. In \cite{zuo112}, a selective relay based cooperative spectrum sensing scheme
was proposed that does not require a separate channel for reporting sensing results. In addition, the proposed scheme can achieve
excellent sensing performance with controllable interference to primary users. These existing works, however, only consider
a simple setting with one pair of secondary users.}

%\vspace{10pt}
\section{System and Spectrum Sensing Models}
\label{SystemModel}

In this section, we describe the system and spectrum sensing models. Specifically, sensing performance in terms of
detection and false alarm probabilities are explicitly described.

\subsection{System Model}
\label{System}

We consider a network setting where $N$ pairs of secondary users opportunistically exploit available frequency bands, which belong a primary network,
for their data transmission. \textbf{Note that the optimization model in \cite{R1} is a special case of our model with only one pair of secondary users.} In particular, we will consider both scenarios in which one or multiple radio channels are exploited by these secondary users. We will design synchronized MAC protocols for both scenarios assuming that each channel can be in idle or busy state for a
predetermined periodic interval, which is referred to as a cycle in this paper.

\begin{figure}[!t] 
\centering
\includegraphics[width=90mm]{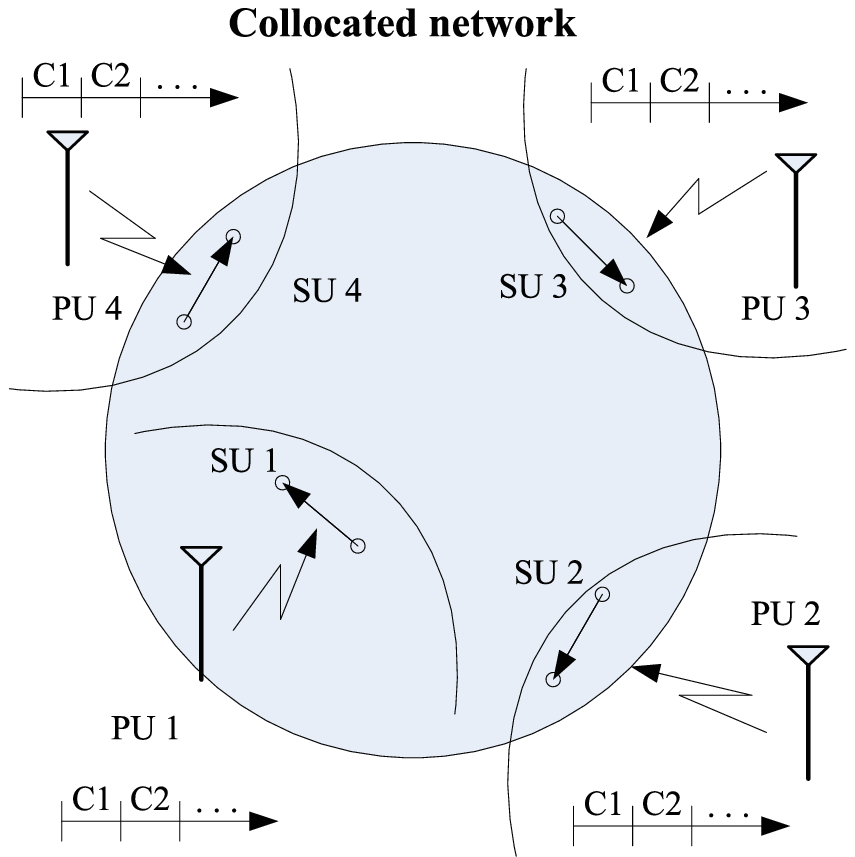}
\caption{Considered network and spectrum sharing model (PU: primary user, SU: secondary user)}
\label{Fig1}
\end{figure}

We further assume that each pair of secondary users can overhear transmissions from other pairs of secondary users (i.e., collocated networks). 
In addition, it is assumed that transmission from each individual pair of secondary users affects one different primary receiver. 
It is straightforward to relax this assumption to the scenario where each pair of secondary users affects more than one primary receiver and/or each primary receiver is affected by more than one  pair of secondary users.
The network setting under investigation is shown in Fig.~\ref{Fig1}. In the following, we will refer to 
pair $i$ of secondary users as secondary link $i$ or flow $i$ interchangeably.

\vspace{0.2cm}
\noindent
\textbf{Remark 1:} \textbf{In practice, secondary users can change their idle/busy status any time (i.e., status changes can occur
in the middle of any cycle). Our assumption on synchronous channel status changes is only needed to estimate the
system throughput. In general, imposing this assumption would not sacrifice the accuracy of our network throughput calculation 
 if primary users maintain their idle/busy status  for 
sufficiently long time on average. This is actually the case for many practical scenarios
such as in TV bands as reported by several recent studies (see \cite{R9} and references therein). In addition, our 
MAC protocols developed under this assumption would result in very few collisions with primary users because
the cycle time is quite small compared to typical active/idle periods of primary users.}

\subsection{Spectrum Sensing}
\label{Ss}

We assume that secondary links rely on a distributed synchronized MAC protocol to share available frequency
channels. Specifically, time is divided into fixed-size cycles and it is assumed that
secondary links can perfectly synchronize with each other (i.e., there is no synchronization error) \cite{Konda08}, 
\cite{R2}. It is assumed that each secondary link performs spectrum sensing at the beginning of each cycle 
and only proceeds to contention with other links to transmit on available channels if its sensing outcomes indicate at least one available channel
(i.e., channels not being used by nearby primary users). For the multiple channel case, we assume that there are $M$ channels
and each secondary transmitter is equipped with $M$ sensors to  sense all channels simultaneously. Detailed MAC protocol 
design will be elaborated in the following sections.

Let $\mathcal{H}_0$ and $\mathcal{H}_1$ denote the events that a particular primary user is idle and active, respectively
(i.e., the underlying channel is available and busy, respectively) in any cycle. In addition, let 
 ${{\mathcal{P}}^{ij}}\left( {{\mathcal{H}_0}} \right)$ and  ${{\mathcal{P}}^{ij}}\left( {{\mathcal{H}_1}} \right) = 1 - {{\mathcal{P}}^{ij}}\left( {{\mathcal{H}_0}} \right)$ be the probabilities that channel $j$ is available and not available at secondary link $i$, respectively.
We assume that secondary users employ an energy detection scheme and let $f_s$ be the sampling frequency used in the
sensing period whose length is $\tau$ for all secondary links. There are two important performance measures,
which are used to quantify the sensing performance, namely detection and false alarm probabilities. In particular,
detection event occurs when a secondary link successfully senses a busy channel and false alarm
represents the situation when a spectrum sensor returns a busy state for an idle channel (i.e., a transmission opportunity
is overlooked).

Assume that transmission signals from primary users are complex-valued PSK signals while the noise at the secondary links is independent and identically distributed circularly symmetric complex Gaussian $\mathcal{CN}\left( {0,{N_0}} \right)$ \cite{R1}. Then, the
detection and false alarm probability for the channel $j$ at secondary link $i$ can be calculated as \cite{R1}
\beqn
\label{eq1}
{{\mathcal{P}}_d^{ij}}\left( {{\varepsilon ^{ij}} ,\tau } \right) = \mathcal{Q}\left( {\left( {\frac{{\varepsilon ^{ij}} }{{{N_0}}} - {\gamma ^{ij}}  - 1} \right)\sqrt {\frac{{\tau {f_s}}}{{2{\gamma ^{ij}}  + 1}}} } \right), 
\eeqn
\beqn
 {{\mathcal{P}}_f^{ij}}\left( {{\varepsilon ^{ij}} ,\tau } \right) = \mathcal{Q}\left( {\left( {\frac{{\varepsilon ^{ij}} }{{{N_0}}} - 1} \right)\sqrt {\tau {f_s}} } \right) \hspace{2.5cm} \nonumber \\ 
 = \mathcal{Q}\left( {\sqrt {2{\gamma ^{ij}}  + 1} {\mathcal{Q}^{ - 1}}\left( {{{\mathcal{P}}_d^{ij}}\left(  {{\varepsilon ^{ij}} ,\tau }  \right)} \right)+\sqrt {\tau {f_s}} {\gamma ^{ij}} } \right),  \label{eq2}
\eeqn
where $i \in \left[ {1,N} \right]$ is the index of a SU link, $j \in \left[ {1,M} \right]$ is the index of a channel, ${\varepsilon ^{ij}} $ is the detection threshold for an energy detector, ${\gamma ^{ij}} $ is the signal-to-noise ratio (SNR) of the PU's signal at the secondary link, $f_s$ is the sampling frequency, $N_0$ is the noise power, $\tau$ is the sensing interval, and $\mathcal{Q}\left( . \right)$ is defined as $\mathcal{Q}\left( x \right) = \left( {1/\sqrt {2\pi } } \right)\int_x^\infty  {\exp \left( { - {t^2}/2} \right)dt}$. In the analysis performed in the following
sections, we assume a homogeneous scenario where sensing performance on different channels is the same for each secondary user.
In this case, we denote these probabilities for secondary user $i$ as  ${\mathcal{P}}_f^{i}$ and ${\mathcal{P}}_d^{i}$ for brevity.

\vspace{0.2cm}
\noindent
\textbf{Remark 2:} \textbf{For simplicity, we do not consider the impact of wireless channel fading in modeling 
the sensing performance in (\ref{eq1}), (\ref{eq2}). This enables us to gain insight into the investigated spectrum sensing and access problem
while keeping the problem sufficiently tractable. Extension of the model to capture wireless fading 
will be considered in our future works. Relevant results published in some recent 
works such as those in \cite{zuo10} would be useful for these further studies}.

\vspace{0.2cm}
\noindent
\textbf{Remark 3:} The analysis performed in the following sections can be easily extended to the case
where each secondary transmitter is equipped with only one spectrum sensor or each secondary transmitter only senses a subset
of all channels in each cycle. Specifically, we will need to adjust the sensing time for some 
spectrum sensing performance requirements. In particular, if only one spectrum sensor is available at each secondary transmitter,
then the required sensing time should be $M$ times larger than the case in which each transmitter has $M$ spectrum sensors.

\section{MAC Design, Analysis and Optimization: Single Channel Case}
\label{SingleChan}

We consider the MAC protocol design, its throughput analysis and optimization
for the single channel case in this section.

\subsection{MAC Protocol Design}
\label{MACDesigna}

% Fig. 2
%\begin{figure}[!t]
%\centering
%\includegraphics[width=80mm]{Sentime}
%\caption{Timing diagram of the proposed single channel MAC protocol.}
%\label{Sentime}
%\end{figure}

\begin{figure}[!t]
\centering
\includegraphics[width=90mm]{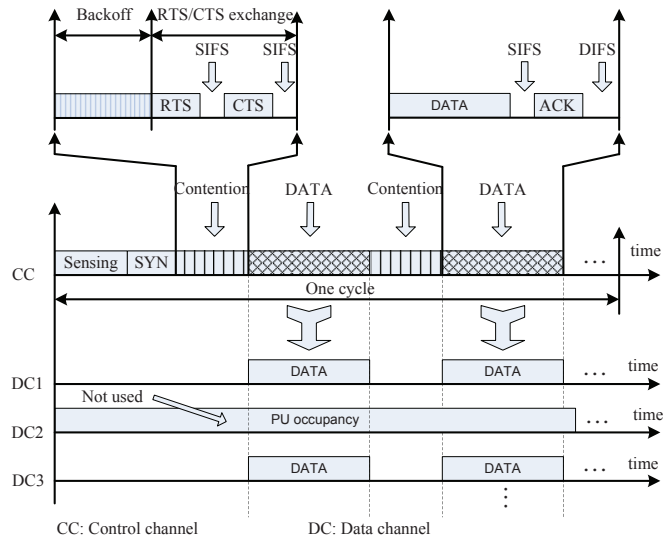}
\caption{Timing diagram of the proposed multi-channel MAC protocol.}
\label{MACoperation}
\end{figure}

We now describe our proposed synchronized MAC for dynamic spectrum sharing among secondary flows.
We assume that each fixed-size cycle of length $T$ is divided into 3 phases,
namely sensing phase, synchronization phase, and data transmission phase. During the sensing 
phase of length $\tau$, all secondary users perform spectrum sensing on the underlying channel.
Then, only secondary links whose sensing outcomes indicate an available channel proceed to the next phase
(they will be called active secondary users/links in the following). In the synchronization phase, 
active secondary users broadcast beacon signals for synchronization purposes. Finally, active secondary users perform contention
and transmit data in the data transmission phase. The timing diagram of one particular cycle is illustrated in Fig.~\ref{MACoperation}.
For this single channel scenario, synchronization, contention, and data transmission occur on the same channel.

We assume that the length of each cycle is sufficiently large so that secondary links can transmit several packets during
the data transmission phase. Indeed, the current 802.22 standard specifies the spectrum evacuation time upon the return
of primary users is 2 seconds, which is a relatively large interval. Therefore, our assumption would be valid for most practical
cognitive systems. During the data transmission phase, we assume that
active secondary links employ a standard contention technique to capture the channel similar to that in the CSMA/CA protocol.
Exponential backoff with minimum contention window $W$ and maximum backoff stage $m$ \cite{R3} is employed in the contention phase.
For brevity, we refer to $W$ simply as contention window in the following.
Specifically, suppose that the current backoff stage of a particular secondary user is $i$ then it starts the contention by choosing a random
backoff time uniformly distributed in the range $[0,2^i W-1]$, $0 \leq i \leq m$. This user then starts decrementing its backoff time counter while carrier sensing transmissions from other secondary links. 

Let $\sigma$ denote a mini-slot interval, each of which corresponds one unit of the
backoff time counter. Upon hearing a transmission from any secondary link, each secondary link will ``freeze"
its backoff time counter and reactivate when the channel is sensed idle again. Otherwise, if the backoff time counter reaches zero,
the underlying secondary link wins the contention. Here, either two-way or four-way handshake with RTS/CST will be employed
to transmit one data packet on the available channel. In the four-way handshake, the transmitter sends 
RTS to the receiver and waits until it successfully receives CTS before sending a data packet. In both handshake schemes,
after sending the data packet the transmitter expects
an acknowledgment (ACK) from the receiver to indicate a successful reception of the packet. Standard small intervals, namely DIFS and SIFS,
are used before backoff time decrements and ACK packet transmission as described in \cite{R3}. We refer to this two-way handshaking technique as a basic access scheme in the following analysis.

\subsection{Throughput Maximization}
\label{TputOpt}

Given the sensing model and proposed MAC protocol, we are interested in finding its optimal configuration to achieve the maximum throughput
subject to protection constraints for primary receivers. Specifically, let $\mathcal{NT}(\tau, W)$ be the normalized total throughput,
which is a function of sensing time $\tau$ and contention window $W$. Suppose that each primary receiver requires that detection probability
achieved by its conflicting primary link $i$ be at least $\overline{P}_d^i$. Then, the throughput maximization problem can be stated
as follows:

\vspace{0.2cm}
\noindent
\textbf{Problem 1:} 
%\vspace{0.1cm}

\begin{equation}
\label{eq3a}% eq3
\begin{array}{l}
 {\mathop {\max }\limits_{\tau ,W}} \quad {\mathcal{NT}} \left( {\tau ,W} \right)  \\ 
 \mbox{s.t.}\,\,\,\, {\mathcal{P}}_d^i\left( {{\varepsilon ^i},\tau } \right) \geq \mathcal{\bar {P}}_d^i, \quad i=1, 2,\cdots, N \\
 \quad \quad 0 < \tau  \le {T},  \quad 0< W \leq W_{\sf max}, \\
 \end{array}\!\!
\end{equation}
where $W_{\sf max}$ is the maximum contention window and recall that $T$ is the cycle interval.
In fact, optimal sensing $\tau$ would allocate sufficient time to protect primary receivers and
optimal contention window would balance between reducing collisions among active secondary links
and limiting protocol overhead. 

\subsection{Throughput Analysis and Optimization}
\label{ThroughputAO}

We perform saturation throughput analysis and solve the optimization problem (\ref{eq3a}) in this subsection.
Throughput analysis for the cognitive radio setting under investigation is more involved compared to
standard MAC protocol throughput analysis (e.g., see \cite{R2}, \cite{R3}) because the number of active secondary 
links participating in the contention in each cycle varies depending on the sensing outcomes. Suppose that 
all secondary links have same packet length. Let  $\Pr \left( {n = {n_0}} \right)$ and 
$\mathcal{T}\left( {\tau ,\phi \left| {n = {n_0}} \right.} \right)$ be the probability that
$n_0$ secondary links participating in the contention and the conditional normalized throughput when $n_0$ secondary links join the channel
 contention,  respectively. Then, the normalized throughput  can be calculated as 
\begin{equation}
\label{ntputa} % ntput
\mathcal{NT} = \sum\limits_{{n_0} = 1}^N {\mathcal{T}\left( {\tau , W\left| {n = {n_0}} \right.} \right) \Pr \left( {n = {n_0}} \right)},
\end{equation}
where recall that $N$ is the number of secondary links, $\tau $ is the sensing time, $W$ is the contention window.
In the following, we show how to calculate $\Pr \left( {n = {n_0}} \right)$ and $\mathcal{T}\left( {\tau ,\phi \left| {n = {n_0}} \right.} \right)$.

\subsubsection{Calculation of $\Pr \left( {n = {n_0}} \right)$} 
\label{CalPrn0}
It is noted that only secondary links whose sensing outcomes in the sensing phase indicate an available channel
proceed to contention in the data transmission phase. There are two scenarios for which
this can happen for a particular secondary link $i$:

\begin{itemize}

\item 
The primary user is not active and no false alarm is generated by the underlying secondary link.

\item
The primary user is active and secondary link $i$ mis-detects its presence.

\end{itemize}
Therefore, secondary link $i$ joins contention in the data transmission phase with probability
\begin{equation}
\label{eq4}
{{\mathcal{P}}_{idle}^i} = \left[ {1 - {{\mathcal{P}}_f^i}\left( {{\varepsilon ^i},\tau } \right)} \right]{{\mathcal{P}}^i}\left( {{\mathcal{H}_0}} \right) + {{\mathcal{P}}_m^i}\left( {{\varepsilon ^i} ,\tau} \right){{\mathcal{P}}^i} \left( {{\mathcal{H}_1}} \right),
\end{equation}
where ${{\mathcal{P}}_m^i}\left( {{\varepsilon ^i} ,\tau} \right)  = 1- {{\mathcal{P}}_d^i}\left( {{\varepsilon ^i} ,\tau} \right)$ is the mis-detection probability.
Otherwise, it will be silent for the whole cycle and waits until the next cycle. This occurs with probability
\beqn \label{Pbusy1ch}
\begin{array}{l}
 \mathcal{P}_{busy}^i = 1 - \mathcal{P}_{idle}^i =  \\ 
 \,\,\,\,\,\,\,\,\,\,\,\,\,\,\,\,\, = \mathcal{P}_f^i\left( {{\varepsilon ^i},\tau } \right){\mathcal{P}^i}\left( {{\mathcal{H}_0}} \right) + \mathcal{P}_d^i\left( {{\varepsilon ^i},\tau } \right){\mathcal{P}^i}\left( {{\mathcal{H}_1}} \right) \\ 
 \end{array}.
\eeqn
We assume that interference of active primary users to the secondary
user is negligible; therefore, a transmission from any secondary link only fails when it collides with transmissions from other secondary links.
Now, let $\mathcal{S}_k$ denote one particular
subset of all secondary links having exactly $n_0$ secondary links. There are $C_N^{{n_0}} = \frac{{N!}}{{{n_0}!(N - {n_0})!}}$ such
sets $\mathcal{S}_k$. The probability of the
 event that $n_0$ secondary links join contention in the data transmission phase can be calculated as
\begin{equation}
\label{eq5}
\Pr \left( {n = {n_0}} \right) = \sum\limits_{k = 1}^{C_N^{{n_0}}} {\prod\limits_{i \in {\mathcal{S}_k}} {{\mathcal{P}}_{idle}^i} \prod\limits_{j \in \mathcal{S}\backslash {\mathcal{S}_k}} {{\mathcal{P}}_{busy}^j} },
\end{equation}
where  $\mathcal{S}$ denotes the set of all $N$ secondary links, and ${\mathcal{S}\backslash {\mathcal{S}_k}}$ is the 
complement of $\mathcal{S}_k$ with  $N-n_0$ secondary links. If all secondary links have the same $SNR_p$ and  the same probabilities ${{\mathcal{P}}^i}\left( {{\mathcal{H}_0}} \right)$ and ${{\mathcal{P}}^i}\left( {{\mathcal{H}_1}} \right)$, then we have
${\mathcal{P}}_{idle}^i = {\mathcal{P}}_{idle}$ and ${\mathcal{P}}_{busy}^i = {\mathcal{P}}_{busy} = 1-{\mathcal{P}}_{idle}$ for all $i$. In this case,
(\ref{eq5}) becomes
\begin{equation}
\label{eq6}
\Pr \left( {n = {n_0}} \right) = C_N^{{n_0}}{\left( {1 - {{\mathcal{P}}_{busy}}}\right)^{{n_0}}}{\left( {{{\mathcal{P}}_{busy}}}  \right)^{N - {n_0}}},
\end{equation}
where all terms in the sum of (\ref{eq5}) become the same.

\vspace{0.2cm}
\noindent
\textbf{Remark 4:} \textbf{In general, interference from active primary users will impact transmissions of
secondary users. However, strong interference from primary users would imply high SNR of sensing signals collected at primary users.
In this high SNR regime, we typically require small sensing time while still satisfactorily protecting primary users. Therefore, for the 
case in which interference from active primary users to secondary users is small, sensing time will have the most significant impact
on the investigated sensing-throughput tradeoff. Therefore, consideration of this setting enables us to gain better 
insight into the underlying problem. Extension to the more general case is possible by explicitly calculating transmission rates
achieved by secondary users as a function of SINR. Due to the space constraint, we will not explore this issue further in this paper.
}

\subsubsection{Calculation of Conditional Throughput}
\label{Con_thoughput}

The conditional throughput can be calculated by using the technique developed by Bianchi in \cite{R3} where
we approximately assume a fixed transmission probability $\phi$ in a generic slot time. 
Specifically, Bianchi shows that this transmission probability can be calculated from
the following two equations \cite{R3}
\beqn \label{phi}
\phi  = \frac{{2\left( {1 - 2p} \right)}}{{\left( {1 - 2p} \right)\left( {W + 1} \right) + Wp\left( {1 - {{\left( {2p} \right)}^m}} \right)}},
\eeqn
\beqn \label{p}
p = 1-\left(1-\phi\right)^{n-1},
\eeqn
where $m$ is the maximum backoff stage, $p$ is the conditional collision probability (i.e., the probability that a collision is observed when a data packet is transmitted on the channel).

Suppose there are $n_0$ secondary links participating in contention in the third phase, the probability of the event that
at least one secondary link transmits its data packet can be written as
\begin{equation}
\label{eq8a} %eq8
{{\mathcal{P}}_t} = 1 - {\left( {1 - \phi } \right)^{{n_0}}}.
\end{equation}
However, the probability that a transmission occurring on the channel is successful given there is at least one secondary
link transmitting can be written as
\begin{equation}
\label{eq9a}% eq9
{{\mathcal{P}}_s} = \frac{{{n_0}\phi {{\left( {1 - \phi } \right)}^{{n_0} - 1}}}}{{{{\mathcal{P}}_t}}}.
\end{equation}
The average duration of a generic slot time can be calculated as 
\begin{equation}
\label{eq10a}% eq10
{{\bar T}_{sd}} = \left( {1 - {{\mathcal{P}}_t}} \right){T_e} + {{\mathcal{P}}_t}{{\mathcal{P}}_s}{T_s} + {{\mathcal{P}}_t}\left( {1 - {{\mathcal{P}}_s}} \right){T_c},
\end{equation}
where $T_e = \sigma$, $T_s$ and $T_c$ represent the duration of an empty slot, the average time the channel is sensed busy due to a successful transmission, and the average time the channel is sensed busy due to a collision, respectively. These quantities can be calculated as \cite{R3}
\\
\textit{For basic mechanism:}
\begin{equation}
\label{eq11a}% eq11
\left\{ \!\!\!{\begin{array}{*{20}{c}}
   {{T_s} = T_s^{1} = H + PS + SIFS + 2PD \!+ \!ACK \!+\! DIFS} \hfill  \\
   {{T_c} = T_c^{1} = H + PS + DIFS + PD} \hfill  \\
   {H = {H_{PHY}} + {H_{MAC}}} \hfill  \\
\end{array}} \right.\!\!\!\!,
\end{equation}
where $H_{PHY}$ and $H_{MAC}$ are the packet headers for physical and MAC layers, $PS$ is the packet size, which is assumed to be fixed in this paper, $PD$ is the propagation delay, $SIFS$ is the length of a short interframe space, $DIFS$ is the length of a distributed interframe space, $ACK$ is the length of an acknowledgment. 
\\
\textit{For RTS/CTS mechanism:}
\beqn \label{TsTc}
\left\{ {\begin{array}{*{20}{c}}
   \begin{array}{l}
 {T_s} = T_s^2 = H + PS + 3SIFS + 2PD + \\ 
 \,\,\,\,\,\,\,\,\,\,\,\,\,\,\,\,\,\,\,\,\,\,\,\,\,\, RTS + CTS + ACK + DIFS \\ 
 \end{array} \hfill  \\
   {{T_c} = T_c^2 = H + DIFS + RTS + PD} \hfill  \\
\end{array}} \right.,
\eeqn
where we abuse notations by letting $RTS$ and $CTS$ represent the length of $RTS$ and $CTS$ control packets, respectively.

Based on these quantities, we can express the conditional normalized throughput as follows:
\begin{equation}
\label{eq13a}%eq13
{\mathcal{T}} \left( {\tau ,\phi \left| {n = {n_0}} \right.} \right) = \left\lfloor {\frac{{{T} - \tau }}{{{{\bar T}_{sd}}}}} \right\rfloor \frac{{{{\mathcal{P}}_s}{{\mathcal{P}}_t}PS}}{{{T}}},
\end{equation}
where $\left\lfloor  .  \right\rfloor $ denotes the floor function and recall that $T$ is the duration of a cycle. Note
that $\left\lfloor {\frac{{{T} - \tau }}{{{{\bar T}_{sd}}}}} \right\rfloor$ denotes the average number of generic slot times
in one particular cycle excluding the sensing phase. Here, we omit the length of the synchronization phase, which is assumed to be negligible.

\subsubsection{Optimal Sensing and MAC Protocol Design}
\label{Optimal_SSa}

Now, we turn to solve the throughput maximization problem formulated in (\ref{eq3a}). Note
that we can calculate the normalized throughput given by (\ref{ntputa}) by using $\Pr \left( {n = {n_0}} \right)$
calculated from (\ref{eq5}) and the conditional throughput calculated from (\ref{eq13a}). It can be
observed that the detection probability ${\mathcal{P}}_d^i\left( {{\varepsilon ^i},\tau } \right)$
in the primary protection constraints ${\mathcal{P}}_d^i\left( {{\varepsilon ^i},\tau } \right) \ge \,\mathcal{\bar {P}}_d^i$
depends on both detection threshold ${\varepsilon ^i}$ and the optimization variable $\tau$. 

We can show that by optimizing the normalized throughput over $\tau$ and $W$ while fixing detection thresholds 
${\varepsilon ^i} = {\varepsilon ^i_0}$ where ${\mathcal{P}}_d^i\left( {{\varepsilon ^i_0},\tau } \right) = \,\mathcal{\bar {P}}_d^i, \: i=1, 2,\cdots , N$, we can achieve almost the maximum throughput gain. The intuition behind this observation can be interpreted as follows.
If we choose ${\varepsilon ^i} < {\varepsilon ^i_0}$ for a given $\tau$, then both ${\mathcal{P}}_d^i\left( {{\varepsilon ^i},\tau }\right)$
and ${\mathcal{P}}_f^i\left( {{\varepsilon ^i},\tau }\right)$ increase compared to the case ${\varepsilon ^i} = {\varepsilon ^i_0}$.
As a result, ${{\mathcal{P}}_{busy}^i}$ given in (\ref{Pbusy1ch}) increases. Moreover, it can be verified that the increase in ${{\mathcal{P}}_{busy}^i}$ will lead to the shift of the probability distribution $\Pr \left( {n = {n_0}} \right)$ to the left. 
Specifically, $\Pr \left( {n = {n_0}} \right)$ given in (\ref{eq5}) increases for small $n_0$ and decreases for large $n_0$ as ${{\mathcal{P}}_{busy}^i}$ increases. 
Fortunately, with appropriate choice of contention window $W$ the conditional throughput ${\mathcal{T}} \left( { \tau ,W\left| {n = n_0} \right.} \right)$ given in (\ref{eq13a}) is quite flat for different $n_0$ (i.e., it only decreases slightly when $n_0$ increases). 
Therefore, the normalized throughput given by (\ref{ntputa}) is almost a constant when we choose ${\varepsilon ^i} < {\varepsilon ^i_0}$.

In the following, we will optimize the normalized throughput over $\tau$ and $W$ while choosing detection thresholds such that ${\mathcal{P}}_d^i\left( {{\varepsilon ^i_0},\tau } \right) = \,\mathcal{\bar {P}}_d^i, \: i=1, 2,\cdots , N$. From these equality constraints and (\ref{eq2}) we have
\beqn \label{Pfoa}%Pfo
 {{\mathcal{P}}_f^i} = \mathcal{Q} \left( \alpha^i + \sqrt {\tau {f_s}} {{\gamma ^i} } \right)
\eeqn
where $\alpha^i =\sqrt {2{\gamma ^i}  + 1} {\mathcal{Q}^{ - 1}} \left( {\bar{\mathcal{P}}_d^i} \right)$.
Hence, the optimization problem (\ref{eq3a}) becomes independent of all detection thresholds ${\varepsilon ^i}, \: i=1,2,\cdots , N$.
Unfortunately, this optimization problem is still a mixed integer program (note that $W$ takes integer values), which is difficult to solve. 
In fact, it can be verified even if we allow $W$ to be a real number, the resulting optimization problem is still not convex
because the objective function is not concave \cite{R6}. Therefore, standard convex optimization techniques cannot be employed
to find the optimal solution for the optimization problem under investigation. 
Therefore, we have to rely on numerical optimization \cite{R7} to find the optimal configuration for the proposed MAC protocol.
Specifically, for a given contention window $W$ we can find the corresponding optimal sensing time $\tau$ as follows:

\vspace{0.2cm}
\noindent
\textbf{Problem 2:} \\
\begin{equation}
\label{eq14}
{\mathop {\max }_{0<\tau \leq T} } \quad {{\mathcal{NT}} \!\!\left( { \tau, W} \right)}  
= \sum\limits_{{n_0} = 1}^N \!\!\! {{\mathcal{T}} \left( { \tau ,W\left| {n = n_0} \right.} \right) \!\Pr \left( {n = {n_0}} \right)}.
\end{equation}

This optimization problem is not convex because its objective function is not concave in general. 
However, we will prove that ${{\mathcal{NT}} \!\!\left( { \tau } \right)}$ is an unimodal function in the range of $\left[0,T\right]$. Specifically, ${{\mathcal{NT}} \!\!\left( { \tau } \right)}$ is monotonically increasing in $\left[0,{\overline{\tau}}\right)$ while it is monotonically decreasing in $\left(\overline{\tau},T\right]$ for some $0 < \overline{\tau} \leq T$. Hence, ${{\mathcal{NT}} \!\!\left( { \overline{\tau} } \right)}$ is the only global maximum in the entire range of $\left[0,T\right]$.
This property is formally stated in the following proposition.

\vspace{0.2cm}
\noindent
\textbf{Proposition 1:} The objective function ${{\mathcal{NT}} \!\!\left( { \tau } \right)}$ of (\ref{eq14}) satisfies
the following properties
\begin{enumerate}
\item %{\beqn \label{eq15}
$\mathop {\lim }\limits_{\tau  \to {T}}   \frac{{\partial {\mathcal{NT}}}}{{\partial \tau }}  < 0$,

\item
$\mathop {\lim }\limits_{\tau  \to 0} \frac{{\partial {\mathcal{NT}}}}{{\partial \tau }} =  + \infty$, %\label{eq17}
%\eeqn}

\item {there is an unique ${\overline{\tau}}$ where ${\overline{\tau}}$ is in the range of $\left[0,T\right]$ such that $\frac{{\partial \mathcal{{NT}}\left( {{\overline{\tau} }} \right)}}{{\partial \tau }} = 0$,}
	\item the objective function ${{\mathcal{NT}} \!\!\left( { \tau } \right)}$ is bounded from above.
\end{enumerate}

\begin{proof} The proof is provided in Appendix A. \end{proof}

We would like to discuss the properties stated in Proposition 1. Properties 1, 2, and 4 imply that there must be at least one $\tau$ in  $\left[0,T\right]$ that maximizes ${\mathcal{{NT}}\left( \tau  \right)}$. The second property implies that indeed 
such an optimal solution is unique. 
Therefore, one can find the globally optimal $(W^*,\tau^*)$ by finding optimal $\tau$ for each $W$ in its feasible range $[1,W_{\sf max}]$. The procedure to find $(W^*,\tau^*)$ can be described in Algorithm 1. Numerical studies reveal that this algorithm has quite low computation time for practical values of $W_{\sf max}$ and $T$.

\begin{algorithm}[h]%\leesize
\caption{\textsc{Optimization of Cognitive MAC Protocol}}
\label{mainalg}
%\algsetup{indent=1.5em}
\begin{algorithmic}[1]

\STATE For each integer value of $W \in [1,W_{\sf max}]$, find the optimal $\tau$ according to (\ref{eq14}), i.e.,
\beqn
\overline{\tau}(W) = {\mathop {\argmax  }_{0<\tau \leq T} } \quad {{\mathcal{NT}} \!\!\left( { \tau, W} \right)}  
\eeqn
\STATE The globally optimal $(W^*,\tau^*)$ can then be found as
\beqn
(W^*,\tau^*) = \argmax_{W, \overline{\tau}(W)}  \quad {{\mathcal{NT}} \!\!\left( { \overline{\tau}(W), W} \right)}.
\eeqn
\end{algorithmic}
\end{algorithm}

\subsection{Some Practical Implementation Issues}

\textbf{
Deployment for the optimal configuration of the proposed MAC protocol can be done as follows.
Each secondary user will need to spend some time to estimate the channel availability probabilities,
channel SNRs, and the number of secondary users sharing the underlying spectrum. When these
system parameters have been estimated, each secondary user can independently calculate the optimal
sensing time and minimum contention window and implement them. Therefore, implementation for optimal MAC protocol can
be performed in a completely distributed manner, which would be very desirable.}

\section{MAC Design, Analysis, and Optimization: Multiple Channel Case}
\label{MultipleChan}

We consider the MAC protocol design, analysis and optimization for the multi-channel case in this section.

\subsection{MAC Protocol Design}
\label{MACDesign}

We propose a synchronized multi-channel MAC protocol for dynamic spectrum sharing in this subsection.
To exploit spectrum holes in this case, we assume that there is one control channel which belongs to the secondary network (i.e., it is always available) and $M$ data channels which can be exploited by secondary users. 
We further assume that each transmitting secondary user employ a reconfigurable transceiver which can be tuned to the control channel or
vacant channels for data transmission easily. In addition, we assume that this transceiver can turn on and off
the carriers on the available or busy channels, respectively (e.g., this can be achieved by the OFDM technology). 

There are still three phases for each cycle as in the single-channel case.
However, in the first phase, namely the sensing phase of length $\tau$, all secondary users simultaneously perform spectrum sensing on all
 $M$ underlying channels. Because the control channel is always available, all secondary users exchange beacon signals to achieve synchronization
in the second phase. Moreover, only active secondary links whose sensing outcomes indicate at least one vacant channel participate in
the third phase (i.e., data transmission phase).  
As a result, the transmitter of the winning link in the contention phase will need to inform its receiver about the available channels. 
Finally, the winning secondary link will transmit data on all vacant channels in the data transmission phase. 
The timing diagram of one particular cycle is illustrated in Fig.~\ref{MACoperation}.

Again, we also assume that the length of each cycle is sufficiently large such that secondary links can transmit several packets on each available channel during the data transmission phase. 
In the data transmission phase, we assume that active secondary links adopt the standard contention technique to capture the channels similar to that employed by the CSMA/CA protocol using exponential backoff and either two-way or four-way handshake as described in Section III.
For the case with two-way handshake, both secondary transmitters and receivers need to perform spectrum sensing.
With four-way handshake, only secondary transmitters need to perform spectrum sensing and
 the RTS message will contain additional information about the available channels on which the receiver will receive data packets.
Also, multiple packets (i.e., one on each available channel) are transmitted by the winning secondary transmitter. Finally, the ACK message
will be sent by the receiver to indicate successfully received packets on the vacant channels.

\subsection{Throughput Maximization}
\label{TputOpt}

In this subsection, we discuss how to find the optimal configuration to maximize the normalized throughput under 
sensing constraints for primary users.
Suppose that each primary receiver requires that detection probability achieved by its conflicting primary link $i$ on
 channel $j$ be at least $\overline{P}_d^{ij}$. Then, the throughput maximization problem can be stated as follows:

\vspace{0.2cm}
\noindent
\textbf{Problem 3:} 
\begin{equation}
\label{eq3}
\begin{array}{l}
 {\mathop {\max }\limits_{\tau ,W}} \quad {\mathcal{NT}} \left( {\tau ,W} \right)  \\ 
 \mbox{s.t.}\,\,\,\, {\mathcal{P}}_d^{ij} \left( {{\varepsilon ^{ij}},\tau } \right) \geq \mathcal{\bar {P}}_d ^{ij}, i\in\left[1, N\right], j\in\left[1, M\right]\\
 \quad \quad 0 < \tau  \le {T},  \quad 0< W \leq W_{\sf max}, \\
 \end{array}\!\!
\end{equation}
where ${\mathcal{P}}_d^{ij}$ is the detection probability for secondary user $i$ on channel $j$,
$W_{\sf max}$ is the maximum contention window and recall that $T$ is the cycle interval.
We will assume that for each secondary user $i$,  
${\mathcal{P}}_d\left( {{\varepsilon ^{ij}},\tau } \right)$ and $\mathcal{\bar {P}}_d ^{ij}$ are the same
for all channel $j$, respectively. This would be valid because sensing performance (i.e., captured in 
${\mathcal{P}}_d\left( {{\varepsilon ^{ij}},\tau } \right)$ and ${\mathcal{P}}_f\left( {{\varepsilon ^{ij}},\tau } \right)$)
depends on detection thresholds $\epsilon^{ij}$ and
the SNR $\gamma^{ij}$, which would be the same for different channels $j$. 
In this case, the optimization problem reduces to that of the same form 
as (\ref{eq3a}) although the normalized throughput ${\mathcal{NT}} \left( {\tau ,W} \right)$ will need to be derived
for this multi-channel case. For brevity, we will drop all channel index $j$ in these quantities whenever possible.

%\vspace{10pt}
\subsection{Throughput Analysis and Optimization}
\label{ThroughputAO}

We analyze the saturation throughput and show how to obtain an optimal solution for \textbf{Problem 3 }.
Again we assume that all secondary links transmit data packets of the same length. 
Let  $\Pr \left( {n = {n_0}} \right)$, $\mathbf{E}\left[l\right]$ and $\mathcal{T}\left( {\tau ,\phi \left| {n = {n_0}} \right.} \right)$ denote the probability that $n_0$ secondary links participating in the contention phase, the average number of vacant channels at the winning SU link, and the conditional normalized throughput when $n_0$ secondary links join the contention,  respectively. 
Then, the normalized throughput  can be calculated as 
\begin{equation}
\label{ntput}
\mathcal{NT} = \sum\limits_{{n_0} = 1}^N {\mathcal{T}\left( {\tau , W\left| {n = {n_0}} \right.} \right) \Pr \left( {n = {n_0}} \right) \frac{{\mathbf{E}\left[ {{l}} \right]}}{M}},
\end{equation}
where recall that $N$ is the number of secondary links, $M$ is the number of channels, $\tau $ is the sensing time, $W$ is the contention window.
Note that this is the average system throughput per channel.
We will calculate $\mathcal{T}\left( {\tau ,\phi \left| {n = {n_0}} \right.} \right)$ using (\ref{eq13a}) for the proposed MAC protocol with four-way handshake and exponential random backoff. In addition, we also show how to calculate $\Pr \left( {n = {n_0}} \right)$ .

\subsubsection{Calculation of $\Pr \left( {n = {n_0}} \right)$ and $\mathbf{E} \left[ l\right]$} 

Recall that only secondary links whose sensing outcomes indicate at least one available channel participate in
contention in the data transmission phase. Again,  as in the single channel case derived in Section \ref{CalPrn0} the
sensing outcome at secondary user $i$ indicates that channel $j$ is available or busy
with probabilities ${\mathcal{P}^{i}_{idle}}$ and ${\mathcal{P}^{i}_{busy}}$, which
are in the same forms with (\ref{Pbusy1ch}) and (\ref{eq4}), respectively (recall that
we have dropped the channel index $j$ in these quantities). 
Now, $\Pr \left( {n = {n_0}} \right)$ can be calculated from these probabilities. 
Recall that secondary link $i$ only joins the contention if its sensing outcomes indicate at least one vacant channel. 
Otherwise, it will be silent for the whole cycle and waits until the next cycle. 
This occurs if its sensing outcomes indicate that all channels are busy.

To gain insight into the optimal structure of the optimal solution while keeping mathematical details sufficiently tractable,
we will consider the homogeneous case in the following where  ${\mathcal{P}^{i}_f}$,  ${\mathcal{P}^{i}_d}$ (therefore,
${\mathcal{P}^{i}_{idle}}$ and ${\mathcal{P}^{i}_{busy}}$) are the same for all secondary users $i$. The obtained
results, however, can extended to the general case even though the corresponding expressions will be more lengthy and tedious.
For the homogeneous system, we will simplify $\mathcal{P}^{i}_{SUidle}$ and $\mathcal{P}^{i}_{SUbusy}$ to $\mathcal{P}_{SUidle}$ and $\mathcal{P}_{SUbusy}$, respectively for brevity. 
Therefore, the probability that a particular channel is indicated as busy or idle by the corresponding spectrum sensor can be written as
\beq \label{Pbus1}
{\mathcal{P}_{busy}} = {\mathcal{P}_f}\mathcal{P}\left( {{\mathcal{H}_0}} \right) + {\mathcal{P}_d}\mathcal{P}\left( {{\mathcal{H}_1}} \right),
\eeq
\beqn \label{Pidl}
{\mathcal{P}_{idle}} = 1 - {\mathcal{P}_{busy}}.
\eeqn
Let $Pr\left(l = l_0\right)$ denote the probability that $l_0$ out of $M$ channels are indicated as available by the spectrum sensors.
Then, this probability can be calculated as
\beqn \label{Pl0}
\Pr \left( {l = {l_0}} \right) = \left( {\begin{array}{*{20}{c}}
   M  \\
   {{l_0}}  \\
\end{array}} \right)\mathcal{P}_{idle}^{{l_0}}\mathcal{P}_{busy}^{M - {l_0}}.
\eeqn
Now, let $\mathcal{P}_{SUidle}$ be the probability that a particular secondary link $i$ participates in the contention (i.e., its spectrum sensors
indicate at least one available channel) and $\mathcal{P}_{SUbusy}$ be the probability that secondary link $i$ is silent (i.e., its spectrum
 sensors indicate that all channels are busy). Then, these probabilities can be calculated as
\beqn \label{PSUbus} 
{\mathcal{P}_{SUbusy}} = \Pr \left( {l = 0} \right) = \mathcal{P}_{busy}^M, 
\eeqn
\beqn \label{PSUidl}
{\mathcal{P}_{SUidle}} = \sum\limits_{{l_0} = 1}^M {\Pr \left( {l = {l_0}} \right) = 1 - \mathcal{P}_{SUbusy}}.
\eeqn
 
Again we assume that a transmission from a particular secondary link only fails if it collides with transmissions from other secondary links.
The probability that $n_0$ secondary links join the contention can be calculated by using (\ref{PSUidl}) and (\ref{PSUbus}) as follows: 
\beqn \label{Pn0}
 \begin{array}{l}
 \Pr \left( {n = {n_0}} \right) = \left( {\begin{array}{*{20}{c}}
   N  \\
   {{n_0}}  \\
\end{array}} \right)\mathcal{P}_{SUidle}^{{n_0}}\mathcal{P}_{SUbusy}^{N - {n_0}} \\ 
 \,\,\,\,\,\,\,\,\,\,\,\,\,\,\,\,\,\,\,\,\,\,\,\,\,\,\,\,\,\,\,\, = \left( {\begin{array}{*{20}{c}}
   N  \\
   {{n_0}}  \\
\end{array}} \right){\left( {1 - \mathcal{P}_{busy}^M} \right)^{{n_0}}}\mathcal{P}_{busy}^{M\left( {N - {n_0}} \right)} \\ 
 \end{array}.
\eeqn
From (\ref{Pl0}), we can calculate the average number of available channels, denoted by the expectation $\mathbf{E}\left[l\right]$, at one 
particular secondary link as
\beqn \label{El}
\begin{array}{l}
 \mathbf{E}[l] = \sum\limits_{{l_0} = 0}^M {{l_0}\Pr \left( {l = {l_0}} \right)}  = \sum\limits_{{l_0} = 0}^M {{l_0}\left( {\begin{array}{*{20}{c}}
   M  \\
   {{l_0}}  \\
\end{array}} \right)\mathcal{P}_{idle}^{{l_0}}\mathcal{P}_{busy}^{M - {l_0}}}  \\ 
 \,\,\,\,\,\,\,\,\,\,\,\, = M{\mathcal{P}_{idle}} = M\left( {1 - {\mathcal{P}_{busy}}} \right) \\ 
 \end{array}.
\eeqn

\subsubsection{Optimal Sensing and MAC Protocol Design}
\label{Optimal_SS}

We now tackle the throughput maximization problem formulated in (\ref{eq3}). In this case, the normalized throughput given by (\ref{ntput}) 
can be calculated by using $\Pr \left( {n = {n_0}} \right)$
in (\ref{Pn0}), the conditional throughput in (\ref{eq13a}), and the average number of available channels in (\ref{El}). 
Similar to the single-channel case, we will optimize the normalized throughput over $\tau$ and $W$ while choosing a detection threshold such that ${\mathcal{P}}_d\left( {{\varepsilon _0},\tau } \right) = \,\mathcal{\bar {P}}_d$. Under these equality constraints, the  
false alarm probability can be written as
\beqn \label{Pfo}
 {{\mathcal{P}}_f} = \mathcal{Q} \left( \alpha + \sqrt {\tau {f_s}} {{\gamma } } \right)
\eeqn
where $\alpha =\sqrt {2{\gamma }  + 1} {\mathcal{Q}^{ - 1}} \left( {\bar{\mathcal{P}}_d} \right)$. 
Hence, \textbf{Problem 3} is independent of detection thresholds.
Again, for a given contention window $W$ we can find the corresponding optimal sensing time $\tau$ 
in the following optimization problem

\vspace{0.1cm}
\noindent
\textbf{Problem 4:} 
\beqn \label{Prob2}
\begin{array}{l}
 \begin{array}{*{20}{c}}
   {\mathop {\max }\limits_\tau  } & {\mathcal{\widetilde{NT}}\left( \tau  \right) \buildrel \Delta \over = NT\left( {\tau ,W} \right)\left| {_{W = \bar W}} \right.}  \\
\end{array} \\ 
 \begin{array}{*{20}{c}}
   {s.t.} & {0 \le \tau  \le }  \\
\end{array}T \\ 
 \end{array}.
\eeqn 

Similar to the single-channel case, we will prove that ${\mathcal{\widetilde{NT}}\left( \tau  \right)}$ is a unimodal function in the range of $\left[0,T\right]$. Therefore, there is a unique global maximum in the entire range of $\left[0,T\right]$. This  is indeed the result
of several properties stated in the following proposition. 

\vspace{0.2cm}
\noindent
\textbf{Proposition 2:} The function ${\mathcal{\widetilde{NT}}\left( \tau  \right)}$ satisfies
the following properties
\begin{enumerate}
	\item {$\mathop {\lim \,}\limits_{\tau  \to 0} \frac{{\partial \mathcal{\widetilde{NT}}\left( \tau  \right)}}{{\partial \tau }} > 0$,}
	\item {$\mathop {\lim \,}\limits_{\tau  \to T} \frac{{\partial \mathcal{\widetilde{NT}}\left( \tau  \right)}}{{\partial \tau }} < 0$,}	
	\item {there is an unique $\overline{\tau}$ where $\overline{\tau}$ is in the range of $\left[0,T\right]$ such that $\frac{{\partial \mathcal{\widetilde{NT}}\left( {\overline{\tau}} \right)}}{{\partial \tau }} = 0$,}
	\item {and the objective function ${\mathcal{\widetilde{NT}}\left( \tau  \right)}$ is bounded from above.}
\end{enumerate}
Therefore, it is a unimodal function in the range of $\left[0,T\right]$

\begin{proof} The proof is provided in Appendix B. \end{proof}

Therefore, given one particular value of $W$ we can find a unique optimal $\overline{\tau}(W)$ for the optimization problem (\ref{Prob2}). Then,
we can find the globally optimal $(W^*,\tau^*)$ by finding optimal $\tau$ for each $W$ in its feasible range $[1,W_{\sf max}]$.
The procedure to find $(W^*,\tau^*)$ is the same as that described in \textbf {Algorithm \ref{mainalg}}. 

%\vspace{10pt}
\section{Numerical Results}
\label{Results}

We present numerical results to illustrate throughput performance of the proposed cognitive MAC protocols.
We take key parameters for the MAC protocols from Table II in \cite{R3}. Other parameters are chosen as follows:
cycle time is $T = 100 ms$; mini-slot (i.e., generic empty slot time) is $\sigma = 20 {\mu} s$;
sampling frequency for spectrum sensing is $f_s = 6 MHz$;  bandwidth of PUs' QPSK signals is $6 MHz$. 
In addition, the exponential backoff mechanism with the maximum backoff stage $m$ is employed to reduce collisions.

\subsection{Performance of Single Channel MAC Protocol}
\label{Results1}

For the results in this section, we choose other parameters of the cognitive network as follows. 
The signal-to-noise ratio of PU signals at secondary links $SNR_p^i$ are chosen randomly in the range $[-15, -20] dB$. 
The target detection probability for secondary links and the probabilities  ${{\mathcal{P}}^i}\left( {{\mathcal{H}_0}} \right)$ are chosen randomly in the intervals $[0.7, 0.9]$ and $[0.7, 0.8]$, respectively. The basic scheme is used as a handshaking mechanism for the MAC protocol.

\begin{figure}[!t]
\centering
\includegraphics[width=95mm]{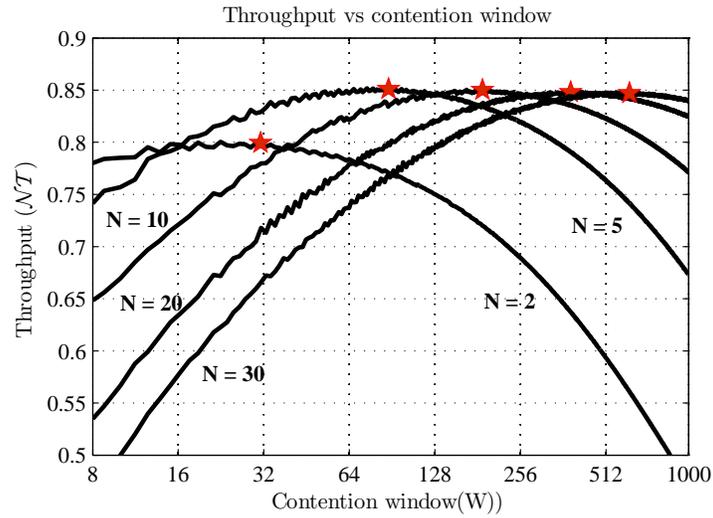}
\caption{Normalized throughput versus contention window $W$ for $\tau = 1 ms$, $m = 3$, different $N$ and basic access mechanism.}
\label{Fig4}
\end{figure}

In Fig.~\ref{Fig4}, we show the normalized throughput $\mathcal{NT}$ versus contention window $W$ for different values of $N$ 
when the sensing time is fixed at $\tau = 1 ms$  and the maximum backoff stage is chosen at $m = 3$ for one particular realization of system parameters. 
The maximum throughput on each curve is indicated by a star symbol. 
This figure indicates that the maximum throughput is achieved at larger $W$ for larger $N$. 
This is expected because larger contention window can alleviate collisions among active secondary for larger number of secondary links.
It is interesting to observe that the maximum throughput can be larger than 0.8 although  ${{\mathcal{P}}^i}\left( {{\mathcal{H}_0}} \right)$ are chosen in the range $[0.7, 0.8]$. 
This is due to a multiuser gain because secondary links are in conflict with difference primary receivers.

\begin{figure}[!t]
\centering
\includegraphics[width=95mm]{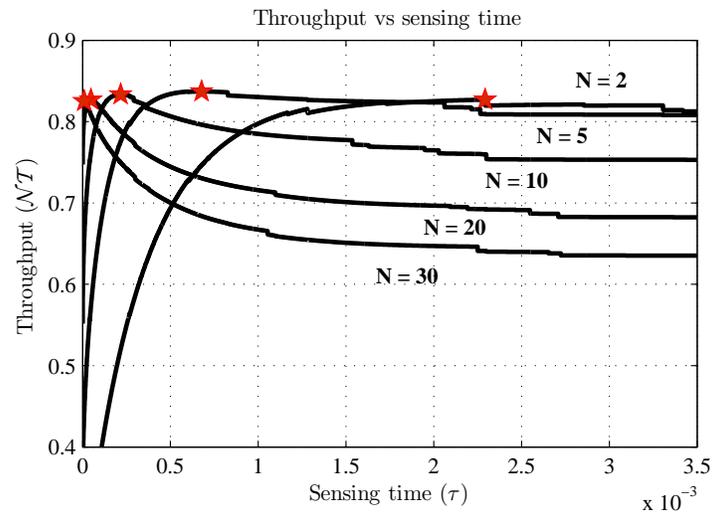}
\caption{Normalized throughput versus the sensing time $\tau$ for $W = 32$, $m = 3$ , different $N$ and basic access mechanism.}
\label{Fig6}
\end{figure}

In Fig.~\ref{Fig6} we present the normalized throughput $\mathcal{NT}$ versus sensing time $\tau$ for a fixed contention window $W=32$, maximum backoff stage $m = 3$, and different number of secondary links $N$. 
The maximum throughput is indicated by a star symbol on each curve. 
This figure confirms that the normalized throughput $\mathcal{NT}$ increases when $\tau$ is small and decreases with large $\tau$ as being proved in Proposition 1. 
%In addition, it can be observed from this figure that the normalized throughput $\mathcal{NT}$ is not a concave
 %function of  $\tau$ for large $N$. 
Moreover, for a fixed contention window the optimal sensing time indeed decreases with the number of secondary links $N$. 
Finally, the multi-user diversity gain can also be observed in this figure.

\begin{figure}[!t]
\centering
\includegraphics[width=95mm]{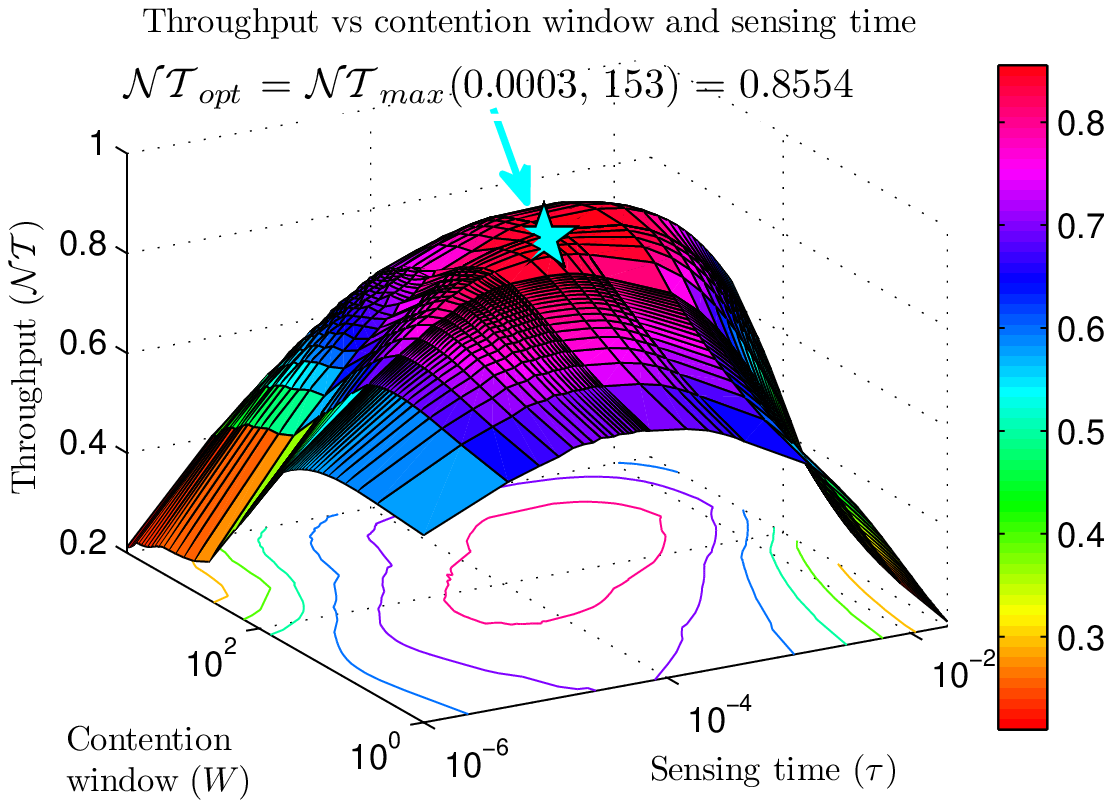}
\caption{Normalized throughput versus sensing time $\tau$ and contention window $W$ for $N = 15$, $m = 4$ and basic access mechanism.}
\label{Fig8}
\end{figure}

To illustrate the joint effects of contention window $W$ and sensing time $\tau$, we show the normalized throughput $\mathcal{NT}$ 
versus $\tau$ and contention window $W$ for $N = 15$ and $m = 4$ in Fig.~\ref{Fig8}.
We show the globally optimal parameters $(\phi^*,\tau^*)$ which maximize the normalized throughput $\mathcal{NT}$  of the proposed cognitive MAC protocol by a star symbol in this figure. 
This figure reveals that the performance gain due to optimal configuration of the proposed MAC protocol is very significant. 
Specifically, while the normalized throughput $\mathcal{NT}$ tends to be less sensitive to the contention window $W$, it decreases 
significantly when the sensing time $\tau$ deviates from the optimal value $\tau^*$.
Therefore, the proposed optimization approach would be very useful in achieving the largest throughput performance for the secondary
network.

\subsection{Performance of Multi-Channel MAC Protocol}
\label{Results2}

In this section, we present numerical results for the proposed multi-channel MAC protocol. Although, we analyze the homogeneous scenario in Section~\ref{MultipleChan} for brevity,  we present simulation results for the heterogeneous settings in this subsection. 
The same parameters for the MAC protocol as in Section \ref{Results1} are used. However, this model covers for the case in which each secondary link has multiple channels. In addition, some key parameters are chosen as follows. The SNRs of the signals from the primary user to secondary link $i$ (i.e., $SNR_p^{ij}$) are randomly chosen in the range of $\left[-15, -20 \right] dB$. The target detection probabilities $\mathcal{\bar P}_d^{ij} $ and the probabilities $\mathcal{P}^{ij} \left(\mathcal{H}_0\right)$ for channel $j$ at secondary link $i$ are randomly chosen in the intervals $\left[0.7, 0.9 \right] $ and $\left[0.7, 0.8 \right] $, respectively. Again the exponential backoff mechanism with the maximum backoff stage $m$ is employed to reduce collisions.

\begin{figure}[!t]
\centering
\includegraphics[width=95mm]{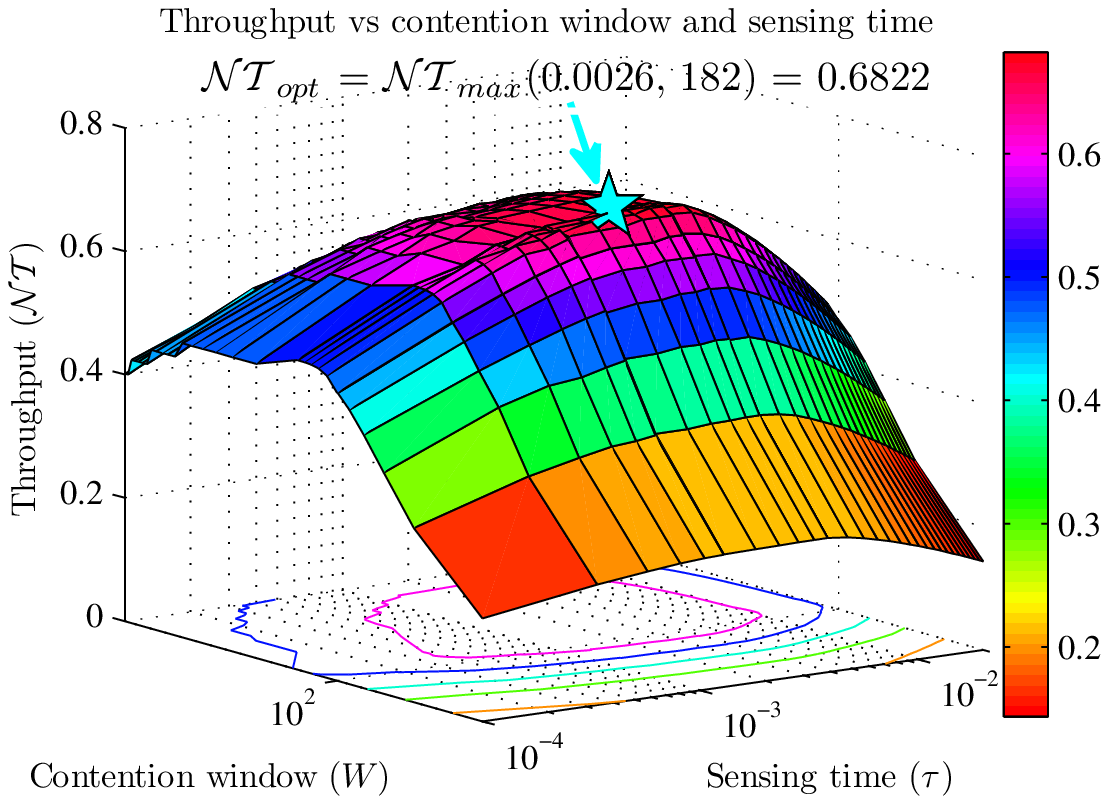}
\caption{Normalized throughput versus sensing time $\tau$ and contention window $W$ for $N = 10$, $m = 4$ , $M = 5$ and basic access mechanism.}
\label{Fig9}
\end{figure}

In Fig.~\ref{Fig9}, we illustrate the normalized throughput $\mathcal{NT}$ versus sensing times $\tau$ and contention windows $W$ for $N = 10$, $M = 5$ and $m = 4$ and the basic access mechanism. 
We show the optimal configuration $\left(\tau^*, W^*\right)$, which maximizes the normalized throughput $\mathcal{NT}$ of the proposed multichannel MAC protocol. 
Again it can be observed that the normalized throughput $\mathcal{NT}$ tends to be less sensitive to the contention window $W$ while it significantly decreases when the sensing time $\tau$ deviates from the optimal sensing time $\tau^*$.

In order to study the joint effect of contention window $W$ and sensing time $\tau$ in greater details, we show the normalized throughput $\mathcal{NT}$ versus $W$ and $\tau$ in Table \ref{table}. 
In this table, we consider both handshaking mechanisms, namely basic access and RTS/CTS access schemes.
Each set of results applies to a particular setting with certain number of secondary links $N$, number of channels $M$ and 
maximum backoff stage $m$. In particular, we will consider two settings, namely $\left(N,M,m\right) = \left(10, 5, 4\right)$ 
and $\left(N,M,m\right) = \left(5, 3, 5\right)$. 
Optimal normalized throughput is indicated by a bold number. It can be confirmed from this table that as
 $\left(\tau, W\right)$ deviate from the optimal $\left(\tau^*, W^*\right)$, the normalized throughput decreases significantly.

This table also demonstrates potential effects of the number of secondary links $N$ on the network throughput and optimal configuration
for the MAC protocols.
In particular, for secondary networks with the small number of secondary links, the probability of collision is lower than that for
networks with the large number of secondary links. 
We consider the two scenarios corresponding to different combinations $\left(N, M, m\right)$. 
The first one which has a smaller number of secondary links $N$ indeed requires smaller contention window 
$W$ and maximum backoff stage $m$ to achieve the maximum throughput. 
Finally, it can be observed that for the same configuration of $\left(N, M, m\right) $, the basic access mechanism slightly outperforms the
RTS/CTS access mechanism, while the RTS/CTS access mechanism can achieve the optimal normalized throughput at lower $W$ compared to
the basic access mechanism.

%\vspace{10pt}
\section{Conclusion}
\label{conclusion} 
In this paper, we have proposed MAC protocols for cognitive radio networks that explicitly take into account spectrum sensing performance.
Specifically, we have derived normalized throughput of the proposed MAC protocols and determined their optimal configuration for throughput 
maximization. These studies have been performed for both single and multiple channel scenarios subject to protection
constraints for primary receivers. Finally, we have presented numerical results to confirm important theoretical findings in the paper and to
show significant performance gains achieved by the optimal configuration for proposed MAC protocols.

%% Fig. 5
%\begin{figure}[!t]
%\centering
%\includegraphics[width=80mm]{Fig5}
%\caption{The normal throughput versus the transmission probability in case of using a RTS/CTS method with $N$ varying and $\tau = 1 ms$. }
%\label{Fig5}
%\end{figure}

%% Fig. 9
%\begin{figure}[!t]
%\centering
%\includegraphics[width=80mm]{Fig9}
%\caption{The normal throughput versus the sensing time and the transmission probability in case of using a RTS/CTS method with $N = 20$.}
%\label{Fig9}
%\end{figure}

\appendices

\section{Proof of Proposition 1}

We start the proof by defining the following quantities: 
$\varphi^j := -\frac{ \left( \alpha^j + \sqrt{\tau f_s}\gamma^j \right)^2} {2}$ 
and ${c_{{n_0}}} := \frac{{{{\mathcal{P}}_s}{{\mathcal{P}}_t}PS}}{{{T}}}$. 
Taking the derivative of $\mathcal{NT}$ versus $\tau$, we have
\begin{equation}
\label{eq16}
\begin{array}{l}
 \frac{{\partial {\mathcal{NT}}}}{{\partial \tau }} = \sum\limits_{{n_0}=1}^N {{c_{{n_0}}}\sum\limits_{k = 1}^{C_{{n_0}}^N} {} }  \\ 
 \left\{\! \!\!\!\begin{array}{l}
 \left( {\frac{{ - 1}}{{{\bar T_{sd}}}}} \right)\prod\limits_{i \in {{\mathcal{S}}_k}} \!\!{{\mathcal{P}}_{idle}^i}\!\! \prod\limits_{j \in {\mathcal{S}}\backslash {{\mathcal{S}}_k}} \!\!{{\mathcal{P}}_{busy}^j}\!\!  + \left\lfloor {\frac{{{T} - \tau }}{{{\bar T_{sd}}}}} \right\rfloor \sqrt {\frac{{{f_s}}}{{8\pi\tau }}}  \times  \\ 
 \left[ \!\!\!\begin{array}{l}
 \sum\limits_{i \in {{\mathcal{S}}_k}} {{\gamma ^i}\exp \left( {{\varphi ^i}} \right) \mathcal{P}^i\left(\mathcal{H}_0 \right) \prod\limits_{l \in {{\mathcal{S}}_k}\backslash i} {{\mathcal{P}}_{idle}^l} \prod\limits_{j \in {\mathcal{S}}\backslash {{\mathcal{S}}_k}} {{\mathcal{P}}_{busy}^j} }  \\ 
  - \sum\limits_{j \in {\mathcal{S}}\backslash {{\mathcal{S}}_k}} {{\gamma ^j}\exp \left( {{\varphi ^j}} \right) \mathcal{P}^j\left(\mathcal{H}_0 \right) \!\!\!\prod\limits_{l \in {\mathcal{S}}\backslash {{\mathcal{S}}_k}\backslash j} \!\!{{\mathcal{P}}_{busy}^l}\!\! \prod\limits_{i \in  {{\mathcal{S}}_k}} \!\!{{\mathcal{P}}_{idle}^i} }  \\ 
 \end{array} \!\!\!\right] \\ 
 \end{array} \!\!\!\!\right\} \\ 
 \end{array}\!\!\!.
\end{equation}
From this we have
\begin{equation}
\label{eq17}
\mathop {\lim }\limits_{\tau  \to {T}}\!\!\!\! \frac{{\partial {\mathcal{NT}}}}{{\partial \tau }} \!\!=\!\! \sum\limits_{{n_0}=1}^N \!\!{{c_{{n_0}}}\sum\limits_{k = 1}^{C_{{n_0}}^N} \!\!{\left(\! {\frac{{ - 1}}{{{\bar T_{sd}}}}} \!\right)\!\!\!\prod\limits_{i \in {{\mathcal{S}}_k}} \!\!\!{{\mathcal{P}}_{idle}^i}\!\!\! \!\!\prod\limits_{j \in {\mathcal{S}}\backslash {{\mathcal{S}}_k}} \!\!\!\!\!{{\mathcal{P}}_{busy}^j} }  < 0}.
\end{equation}
Now, let us define the following quantity 
\begin{equation}
\label{eq18}
{K_\tau } \!\!:= \sum\limits_{{n_0}=1}^N \!{{c_{{n_0}}}\!\sum\limits_{k = 1}^{C_{{n_0}}^N}\!\! {\left[\!\!\! \begin{array}{l}
 \sum\limits_{i \in {{\mathcal{S}}_k}} \!\!{{\gamma ^i}\exp \left( {{\varphi ^i}} \right)\!\! \mathcal{P}^i\!\left(\mathcal{H}_0 \right) \!\!\!\!\!\prod\limits_{l \in {{\mathcal{S}}_k}\backslash i} \!\!\!\!{{\mathcal{P}}_{idle}^l} \!\!\!\!\prod\limits_{j \in {\mathcal{S}}\backslash {{\mathcal{S}}_k}} \!\!\!\! {{\mathcal{P}}_{busy}^j}  - }  \\ 
 \sum\limits_{j \in {\mathcal{S}}\backslash {{\mathcal{S}}_k}}\!\! \!\!\!\!{{\gamma ^j}\exp \left( {{\varphi ^j}} \right) \!\!\mathcal{P}^j\!\left(\mathcal{H}_0 \right) \!\!\!\!\!\!\!\!\prod\limits_{l \in {\mathcal{S}}\backslash {{\mathcal{S}}_k}\backslash j} \!\!\!\!\!\!{{\mathcal{P}}_{busy}^l} \!\!\!\prod\limits_{i \in  {{\mathcal{S}}_k}} \!\!\!{{\mathcal{P}}_{idle}^i} }  \\ 
 \end{array} \!\!\!\right]} }.
\end{equation}

Then, it can be shown that ${K_\tau }>0$ as being explained in the following. 
First, it can be verified that the term $c_{n_0} $ is almost a constant for different $n_0$. Therefore, to highlight intuition behind the 
underlying property (i.e., ${K_\tau }>0$), we substitute $K = c_{n_0}$ into the above equation. Then, ${K_\tau }$ in (\ref{eq18}) reduces to
\beqn \label{Ktau1ch1}
{K_\tau } = {K_a}\sum\limits_{{n_0} = 1}^N {C_{{n_0}}^N\left( {{n_0}\mathcal{P}_{idle}^{{n_0} - 1}\mathcal{P}_{busy}^{N - {n_0}}-\left( {N - {n_0}} \right)\mathcal{P}_{busy}^{N - {n_0} - 1}} \right)} ,
\eeqn
where ${K_a} = K\gamma \exp \left( \varphi  \right)\mathcal{P}\left( {{\mathcal{H}_0}} \right)$. Let define the following quantities $x =  \mathcal{P}_{busy}$, $x \in {R_x} \buildrel \Delta \over = \left[ {{\mathcal{P}_d}\mathcal{P}\left( {{\mathcal{H}_1}} \right),\mathcal{P}\left( {{\mathcal{H}_0}} \right) + {\mathcal{P}_d}\mathcal{P}\left( {{\mathcal{H}_1}} \right)} \right]$. After some manipulations, we have
\beqn \label{Ktau1ch}
{K_\tau } = {K_a}\sum\limits_{{n_0} = 1}^N {f\left( x \right)\left( {\frac{{{n_0}}}{{x\left( {1 - x} \right)}} - \frac{N}{x}} \right)} ,
\eeqn
where $f\left( x \right) = C_{{n_0}}^N{\left( {1 - x} \right)^{{n_0}}}{x^{N - {n_0}}}$ is the binomial mass function \cite{Kris06} with $p = 1 - x$ and $q = x$. Because the total probabilities and the mean of this binomial distribution are 1 and $Np = N\left(1 - x\right)$, respectively, we have 
\beqn \label{bioCDF}
\sum\limits_{{n_0} = 0}^N {f\left( x \right)}  = 1,
\eeqn
\beqn \label{biomean}
\sum\limits_{{n_0} = 0}^N {{n_0}f\left( x \right)}  = N\left( {1 - x} \right).
\eeqn
It can be observed that in (\ref{Ktau1ch}), the element corresponding to $n_0 =0$ is missing.  Apply the results in
 (\ref{bioCDF}) and (\ref{biomean}) to (\ref{Ktau1ch}) we have
\beqn \label{Ktau1ch2}
{K_\tau } = {K_a} N x^{N-1} > 0,  \: \forall x.
\eeqn
Therefore, we have
\begin{equation}
\label{eq19}
\mathop {\lim }\limits_{\tau  \to 0} \frac{{\partial {\mathcal{NT}}}}{{\partial \tau }} =  + \infty.
\end{equation}
Hence, we have completed the proof for first two properties of Proposition 1.

In order to prove the third property, let us find the solution of $\frac{{\partial {\mathcal{NT}}}}{{\partial \tau }} = 0$. 
After some simple manipulations and using the properties of the binomial distribution, this equation reduces to
\beqn \label{hg1ch}
h\left(\tau \right) = g\left(\tau \right),
\eeqn
where 
\beq
g\left( \tau  \right) = {\left( {\alpha  + \gamma \sqrt {{f_s}\tau } } \right)^2},
\eeq
and
\beq
 h\left( \tau  \right) = 2\log \left( {\mathcal{P}\left( {{\mathcal{H}_0}} \right)\gamma \sqrt {\frac{{{f_s}}}{{8\pi }}} \frac{{T - \tau }}{{\sqrt \tau  }}} \right) + {h_1}\left( x \right)
 \eeq 
where 
${h_1}\left( x \right) = 2\log \frac{{{K_\tau }/{K_a}}}{{\sum\limits_{{n_0} = 1}^N {\left( {\begin{array}{*{20}{c}}
   N  \\
   {{n_0}}  \\  
\end{array}} \right)f\left( x \right)} }} = 2\log \frac{{N{x^{N - 1}}}}{{1 - {x^N}}}$. 

To prove the third property, we will show that $h\left(\tau \right)$ intersects $g\left(\tau \right)$ only once.
We first state one important property of $h\left(\tau \right)$ in the following lemma.

\vspace{0.2cm}
\noindent
\textbf{Lemma 1:} $h\left(\tau \right)$ is an decreasing function.

\begin{proof}
Taking the first derivative of $h(.)$, we have 
\beqn \label{partialh1ch} 
\frac{{\partial h}}{{\partial \tau }} = \frac{{ - 1}}{\tau } - \frac{2}{{T - \tau }} + \frac{{\partial {h_1}}}{{\partial x}}\frac{{\partial x}}{{\partial \tau }}. 
\eeqn
We now derive $\frac{{\partial x}}{{\partial \tau }}$ and $\frac{{\partial {h_1}}}{{\partial x}}$  as follows:
\beqn \label{partialx1ch} 
\frac{{\partial x}}{{\partial \tau }} =  - \mathcal{P}\left( {{\mathcal{H}_0}} \right)\gamma \sqrt {\frac{{{f_s}}}{{8\pi \tau }}} \exp \left( { - \frac{{{{\left( {\alpha  + \gamma \sqrt {{f_s}\tau } } \right)}^2}}}{2}} \right) < 0,
\eeqn
\beqn \label{partialh11ch} 
\frac{{\partial {h_1}}}{{\partial x}} = 2 \frac{{N - 1 + {x^N}}}{{x\left( {1 - {x^N}} \right)}}>0.
\eeqn
Hence, $\frac{{\partial {h_1}}}{{\partial x}}\frac{{\partial x}}{{\partial \tau }}<0$. Using this result in (\ref{partialh1ch}), we have $\frac{{\partial h}}{{\partial \tau }} < 0$. Therefore, we can conclude that $h\left(\tau \right)$ is monotonically decreasing.
\end{proof}

We now consider function $g\left(\tau\right)$. Take the derivative of $g\left(\tau\right)$, we have
\beq \label{dgfun}
\frac{{\partial g}}{{\partial \tau }} = \left( \alpha + \gamma\sqrt{f_s \tau} \right) \frac{ \gamma\sqrt{f_s} }{\sqrt{\tau}}.
\eeq
Therefore, the monotonicity property of $g\left(\tau\right)$ only depends on $y = \alpha + \gamma\sqrt{f_s \tau}$.
Properties 1 and 2 imply that there must be at least one intersection between $h\left(\tau \right)$ and $g\left(\tau\right)$.
We now prove that there is indeed a unique intersection. To proceed, we consider two different regions for $\tau$ as follows:

${\mathbf{\Omega}_1} = \left\{ {\tau \left| {\alpha  + \gamma \sqrt {{f_s}\tau }  < 0,\,  \tau  \le T} \right.} \right\}
= \left\{  0< \tau < \frac{\alpha^2 }{\gamma^2 f_s} \right\}$

and 

${\mathbf{\Omega}_2} = \left\{ {\tau \left| {\alpha  + \gamma \sqrt {{f_s}\tau }  \geq 0,\,  \tau  \le T} \right.} \right\}
= \left\{  \frac{\alpha^2 }{\gamma^2 f_s} \leq \tau \leq T \right\}$.

From the definitions of these two regions, we have $g\left(\tau\right)$ decreases in ${\mathbf{\Omega}_1}$ and increases in ${\mathbf{\Omega}_2}$.
To show that there is a unique intersection between $h\left(\tau \right)$ and $g\left(\tau\right)$, we prove the following.

\vspace{0.2cm}
\noindent
\textbf{Lemma 2:} The following statements are correct:
\bnu

\item
If there are intersections between $h\left(\tau \right)$ and $g\left(\tau\right)$ in ${\mathbf{\Omega}_2}$ then it is the only intersection in this region and there is no
intersection in ${\mathbf{\Omega}_1}$.

\item If there are intersections between $h\left(\tau \right)$ and $g\left(\tau\right)$ in ${\mathbf{\Omega}_1}$ then it is the only intersection in this region and there is no
intersection in ${\mathbf{\Omega}_2}$.

\enu

\begin{proof}
We prove the first statement now. Recall that $g\left(\tau\right)$ monotonically increases in ${\mathbf{\Omega}_2}$; therefore, 
$g\left(\tau\right)$ and $h\left(\tau\right)$ can intersect at most once in this region (because $h\left(\tau\right)$ decreases). 
In addition, $g\left(\tau\right)$ and $h\left(\tau\right)$ cannot intersection in ${\mathbf{\Omega}_1}$ for this case if
we can prove that $\frac{{\partial h}}{{\partial \tau }} < \frac{{\partial g}}{{\partial \tau }}$. This is because both
functions decrease in ${\mathbf{\Omega}_1}$. We will prove that $\frac{{\partial h}}{{\partial \tau }} < \frac{{\partial g}}{{\partial \tau }}$
in lemma 3 after this proof.

We now prove the second statement of lemma 2. Recall that we have $\frac{{\partial h}}{{\partial \tau }} < \frac{{\partial g}}{{\partial \tau }}$.
Therefore, there is at most one intersection between $g\left(\tau\right)$ and $h\left(\tau\right)$ in ${\mathbf{\Omega}_1}$. In addition,
it is clear that there cannot be any intersection between these two functions in ${\mathbf{\Omega}_2}$ for this case.
\end{proof}

\vspace{0.2cm}
\noindent
\textbf{Lemma 3:} We have $\frac{{\partial h}}{{\partial \tau }} < \frac{{\partial g}}{{\partial \tau }}$.

\begin{proof} 
From (\ref{partialh1ch}), we can see that lemma 3 holds if we can prove the following stronger result
\beqn \label{partialh1g1ch}
\frac{-1}{\tau}+\frac{{\partial h_1}}{{\partial \tau }} < \frac{{\partial g}}{{\partial \tau }},
\eeqn
where $\frac{{\partial {h_1}}}{{\partial \tau }} = \frac{{\partial {h_1}}}{{\partial x}}\frac{{\partial x}}{{\partial \tau }}$, $\frac{{\partial x}}{{\partial \tau }}$ is derived in (\ref{partialx1ch}), $\frac{{\partial {h_1}}}{{\partial x}}$ is derived in (\ref{partialh11ch})
and  $\frac{{\partial g}}{{\partial \tau }}$ is given in (\ref{dgfun}).

%Let us define the following quantity
%\beq
%
%\eeq

To prove (\ref{partialh1g1ch}), we will prove the following
\beqn \label{partialh1g11chalter}
 - \frac{1}{\tau} + 
 \frac{{y\mathcal{P}\left( {{\mathcal{H}_0}} \right)\gamma \sqrt {\frac{{{f_s}}}{\tau }} }}{{\mathcal{P}\left( {{\mathcal{H}_0}} \right) + \sqrt {2\pi } \mathcal{P}\left( {{\mathcal{H}_1}} \right)\left( {1 - {\mathcal{\bar P}_d}} \right)\left( { - y} \right)\exp \left( {\frac{{{y^2}}}{2}} \right)}} 
 < \frac{{\partial g}} {{\partial \tau }},
\eeqn
where $y = \left(\alpha + \gamma \sqrt{f_s \tau} < 0\right)$.
Then, we show that
\beq \label{eq50}
 \frac{{\partial h_1}}{{\partial \tau }} < \frac{{y\mathcal{P}\left( {{\mathcal{H}_0}} \right)\gamma \sqrt {\frac{{{f_s}}}{\tau }} }}{{\mathcal{P}\left( {{\mathcal{H}_0}} \right) + \sqrt {2\pi } \mathcal{P}\left( {{\mathcal{H}_1}} \right)\left( {1 - {\mathcal{\bar P}_d}} \right)\left( { - y} \right)\exp \left( {\frac{{{y^2}}}{2}} \right)}}.
\eeq
Therefore, the result in (\ref{partialh1g1ch}) will hold. Let us prove (\ref{eq50}) first.
First, let us prove the following 
\beqn \label{h2lower1cha}
\frac{{\partial {h_1}}}{{\partial x}} > \frac{2}{{1 - x}}.
\eeqn
Using the result in $\frac{{\partial {h_1}}}{{\partial x}}$ from (\ref{partialh11ch}), (\ref{h2lower1cha}) is equivalent to
\beqn \label{h2lower1ch}
2\frac{{N - 1 + {x^N}}}{{x\left( {1 - {x^N}} \right)}} > \frac{2}{{1 - x}},
\eeqn
After some manipulations, we get
\beqn \label{h2lower1cha1}
\left( {1 - x} \right)\left( {N - 1 - \left( {x + {x^2} +  \cdots  + {x^{N - 1}}} \right)} \right) > 0.
\eeqn
It can be observed that $0<x<1$ and  $0<x^i<1$, $i \in \left[1,N-1\right]$. So $N-1-\left(x+x^2+\cdots+x^{\left(N-1\right)}\right) >0$; hence (\ref{h2lower1cha1}) holds. Therefore, we have completed the proof for (\ref{h2lower1cha}).

We now show that the following inequality holds
\beqn \label{onex1ch}
\frac{2}{{1 - x}} > \frac{{2\sqrt {2\pi } \left( { - y} \right)\exp \left( {\frac{{{y^2}}}{2}} \right)}}{{\mathcal{P}\left( {{\mathcal{H}_0}} \right) + \sqrt {2\pi } \mathcal{P}\left( {{\mathcal{H}_1}} \right)\left( {1 - {\mathcal{\bar P}_d}} \right)\left( { - y} \right)\exp \left( {\frac{{{y^2}}}{2}} \right)}}.
\eeqn
This can be proved as follows.
In \cite{Geor07}, it has been shown that $Q\left(t\right)$ with $t > 0$ satisfies  
\beqn \label{Qfunction}
\frac{1}{{Q\left( t \right)}} > \sqrt {2\pi } t\exp \left( {\frac{{{t^2}}}{2}} \right).
\eeqn
Apply this result to $\mathcal{P}_f = Q\left(y\right) = 1 - Q\left(-y\right)$ with $y = \left( {\alpha  + \gamma \sqrt {{f_s}\tau } } \right) < 0$ 
we have
\beqn \label{Qfunction1ch}
\frac{1}{{1 - {P_f}}} > \sqrt {2\pi } \left( { - y} \right)\exp \left( {\frac{{{y^2}}}{2}} \right).
\eeqn
After some manipulations, we obtain
\beqn \label{Pf1ch}
{\mathcal{P}_f} > 1 - \frac{1}{{\sqrt {2\pi } \left( { - y} \right)\exp \left( {\frac{{{y^2}}}{2}} \right)}}.
\eeqn
Recall that we have defined $x = {\mathcal{P}_f}\mathcal{P}\left( {{\mathcal{H}_0}} \right) + {\mathcal{\bar P}_d}\mathcal{P}\left( {{\mathcal{H}_1}} \right)$. Using the result in (\ref{Pf1ch}), we can obtain the  lower bound of $\frac{2}{1-x}$ given in (\ref{onex1ch}).
Using the results in (\ref{h2lower1cha}) and (\ref{onex1ch}), and the fact that $\frac{{\partial x}}{{\partial \tau }} < 0$, we 
finally complete the proof for (\ref{eq50}).

%\beqn \label{onex1ch}
%\frac{2}{{1 - x}} > \frac{{2\sqrt {2\pi } \left( { - y} \right)\exp \left( {\frac{{{y^2}}}{2}} \right)}}{{\mathcal{P}\left( {{\mathcal{H}_0}} \right) + \sqrt {2\pi } \mathcal{P}\left( {{\mathcal{H}_1}} \right)\left( {1 - {\mathcal{\bar P}_d}} \right)\left( { - y} \right)\exp \left( {\frac{{{y^2}}}{2}} \right)}}.
%\eeqn
%\beqn \label{partialh1tau1ch}
%\frac{{\partial {h_1}}}{{\partial x}}\frac{{\partial x}}{{\partial \tau }} < \frac{{y\mathcal{P}\left( {{\mathcal{H}_0}} \right)\gamma \sqrt {\frac{{{f_s}}}{\tau }} }}{{\mathcal{P}\left( {{\mathcal{H}_0}} \right) + \sqrt {2\pi } \mathcal{P}\left( {{\mathcal{H}_1}} \right)\left( {1 - {\mathcal{\bar P}_d}} \right)\left( { - y} \right)\exp \left( {\frac{{{y^2}}}{2}} \right)}}.
%\eeqn

To complete the proof of the lemma, we need to prove that (\ref{partialh1g11chalter}) holds.
Substitute $\frac{{\partial g}}{{\partial \tau }}$ from (\ref{dgfun}) to (\ref{partialh1g11chalter}) and make some further manipulations, we have
\beqn \label{finalproof1ch}
\frac{1}{{ - y\left( {y - \alpha } \right)}} > 1 - \frac{{y\mathcal{P}\left( {{\mathcal{H}_0}} \right)\gamma \sqrt {\frac{{{f_s}}}{\tau }} }}{{\mathcal{P}\left( {{\mathcal{H}_0}} \right) + \sqrt {2\pi } \mathcal{P}\left( {{\mathcal{H}_1}} \right)\left( {1 - {\mathcal{\bar P}_d}} \right)\left( { - y} \right)\exp \left( {\frac{{{y^2}}}{2}} \right)}}.
\eeqn
Let us consider the LHS of (\ref{finalproof1ch}).  We have $0<y-\alpha={\gamma \sqrt {{f_s}\tau } }<-\alpha$; therefore, we have
 $0<-y<-\alpha$. Apply the Cauchy–Schwarz inequality to $-y$ and $y - \alpha$, we have the following
\beqn \label{leftside1ch} 
0 <- y\left( {y - \alpha } \right) \leq {\left( {\frac{{ - y + y - \alpha }}{2}} \right)^2} = \frac{{{\alpha ^2}}}{4}.
\eeqn
Hence
\beqn \label{leftside1ch1} 
\frac{1}{{ - y\left( {y - \alpha } \right)}} \ge \frac{4}{{{\alpha ^2}}} = \frac{4}{{\left( {2\gamma  + 1} \right){{\left( {{Q^{ - 1}}\left( {{\mathcal{\bar P}_d}} \right)} \right)}^2}}} > 1.
\eeqn
It can be observed that the RHS of (\ref{finalproof1ch}) is less than 1. Therefore, (\ref{finalproof1ch}) holds, which
implies that (\ref{partialh1g11chalter}) and (\ref{partialh1g1ch}) also hold. 
\end{proof}

Finally, the last property holds because because $\Pr \left( {n = {n_0}} \right)<1$ and conditional throughput are all bounded from above. 
Therefore, we have completed the proof of Proposition 1.

\section{Proof of Proposition 2}

To prove the properties stated in  Proposition 2, we first find the derivative of ${\mathcal{\widetilde{NT}}\left( \tau  \right)}$. 
Again, it can be verified that $\frac{{{\mathcal{P}_t}{\mathcal{P}_s}PS}}{{{T}}}$ is almost a constant for different $n_0$. To demonstrate 
the proof for the proposition, we substitute this term as a constant value, denoted as $\mathcal{K}$, in the throughput formula. In addition, for large $T$, $\left\lfloor {\frac{{{T} - \tau }}{{{{\bar T}_{sd}}}}} \right\rfloor$ is very close to ${\frac{{{T} - \tau }}{{{{\bar T}_{sd}}}}}$. Therefore,  $\mathcal{\widetilde{NT}}$
can be accurately approximated as
\beqn \label{NT_tau}
\mathcal{\widetilde{NT}} \!\!\left( \tau  \right)\!\! =\!\! \sum\limits_{{n_0} = 1}^N {\!\!\mathcal{K}\!\!\left( \!\!\!\!{\begin{array}{*{20}{c}}
   N  \\
   {{n_0}}  \\
\end{array}} \!\!\!\!\right)\!\!\left( {T - \tau } \right)\!\!{{\left( {1 - {x^M}} \right)}^{{n_0}}}{x^{M\left( {N - {n_0}} \right)}}\left( {1 - x} \right)},
\eeqn
where
$\mathcal{K} = \frac{{{P_t}{P_s}PS}}{{{T}}}$, and $x = P_{busy}$. Now, let us define the following function
\beqn \label{fx}
f'\left( x \right) = {\left( {1 - {x^M}} \right)^{{n_0}}}{x^{M\left( {N - {n_0}} \right)}}\left( {1 - x} \right).
\eeqn
Then, we have
%\beqn \label{partialx}
%\frac{{\partial x}}{{\partial \tau }} =  - P\left( {{H_0}} \right)\gamma \sqrt {\frac{{{f_s}}}{{8\pi \tau }}} \exp \left( { - \frac{{{{\left( {\alpha  + \gamma \sqrt {{f_s}\tau } } \right)}^2}}}{2}} \right) < 0,
%\eeqn
\beqn \label{partialfx}
\frac{{\partial f'}}{{\partial x}} = f'\left( x \right)\left[ {\frac{{ - 1}}{{1 - x}} - \frac{{M{n_0}}}{{1 - {x^M}}}{x^{M - 1}} + \frac{{M\left( {N - {n_0}} \right)}}{x}} \right],
\eeqn
and $\frac{{\partial x}}{{\partial \tau }} $ is the same as (\ref{partialx1ch}).
Hence, the first derivation of ${\mathcal{\widetilde{NT}}\left( \tau  \right)}$ can be written as
\beqn \label{deriNT_tau} 
\begin{array}{l}
 \frac{{\partial \mathcal{\widetilde{NT}}\left( \tau  \right)}}{{\partial \tau }} = \sum\limits_{{n_0} = 1}^N {\mathcal{K}\left( {\begin{array}{*{20}{c}}
   N  \\
   {{n_0}}  \\
\end{array}} \right)\left[ { - f'\left( x \right) + \left( {T - \tau } \right)\frac{{\partial f'}}{{\partial x}}\frac{{\partial x}}{{\partial \tau }}} \right]}  \\ 
 \,\,\,\,\,\,\,\,\,\,\,\,\,\,\,\,\,\,\,\,\, = \sum\limits_{{n_0} = 1}^N {\mathcal{K}\left( {\begin{array}{*{20}{c}}
   N  \\
   {{n_0}}  \\
\end{array}} \right)f'\left( x \right)}  \times  \\ 
 \,\,\,\,\,\,\,\,\,\,\,\,\,\,\,\,\,\,\,\left[ \begin{array}{l}
 \left( {T - \tau } \right)\left[ {\frac{1}{{1 - x}} + \frac{{M{n_0}}}{{1 - {x^M}}}{x^{M - 1}} - \frac{{M\left( {N - {n_0}} \right)}}{x}} \right] \\ 
  \times P\left( {{H_0}} \right)\gamma \sqrt {\frac{{{f_s}}}{{8\pi \tau }}} \exp \left( { - \frac{{{{\left( {\alpha  + \gamma \sqrt {{f_s}\tau } } \right)}^2}}}{2}} \right) - 1 \\ 
 \end{array} \right] \\ 
 \end{array}.
\eeqn 
From (\ref{Pbus1}), the range of $x$, namely $\mathbb{R}_x$ can be expressed as $\left[ {{\mathcal{P}_d}\mathcal{P}\left( {{\mathcal{H}_1}} \right),\,\mathcal{P}\left( {{\mathcal{H}_0}} \right) + {\mathcal{P}_d}\mathcal{P}\left( {{\mathcal{H}_1}} \right)} \right]$.
Now, it can be observed that
\beq
\mathop {\lim \,}\limits_{\tau  \to T} \frac{{\partial \mathcal{\widetilde{NT}}\left( \tau  \right)}}{{\partial \tau }} =  - \sum\limits_{{n_0} = 1}^N {\mathcal{K}\left( {\begin{array}{*{20}{c}}
   N  \\
   {{n_0}}  \\
\end{array}} \right)f'\left( x \right)}  < 0.
\eeq
Therefore, the second property of Proposition 2 holds.

%Fig3
%\begin{figure}[t]
%\centering
%\includegraphics[width=80mm]{Ktau}
%\caption{$K_{\tau}$ vs x.}
%\label{Ktau}
%\end{figure}

Now, let us define the following quantity
\beqn \label{Kcons}
{\mathcal{K'}_\tau } \!=\! \sum\limits_{{n_0} = 1}^N {\!\!\!\left(\!\!\!\! {\begin{array}{*{20}{c}}
   N  \\
   {{n_0}}  \\
\end{array}} \!\!\!\!\right)\!\!f'\left( x \right)\left[ {\frac{1}{{1 - x}} \!+\! \frac{{M{n_0}}}{{1 - {x^M}}}{x^{M - 1}} \!- \!\frac{{M\!\left( {N \!- {n_0}} \right)}}{x}} \right]} .
\eeqn
Then, it can be seen that $\mathop {\lim \,}\limits_{\tau  \to 0} \frac{{\partial \mathcal{\widetilde{NT}}\left( \tau  \right)}}{{\partial \tau }} =  + \infty  > 0$ if ${\mathcal{K'}_\tau } > 0,\,\forall M,\,N,\,x \in {\mathbb{R}_x}$. This last property is stated and proved in the following lemma.

\vspace{0.2cm}
\noindent
\textbf{Lemma 4:}
\label{Prop3}
${\mathcal{K'}_\tau } > 0,\,\forall M,\,N,\,x \in {\mathbb{R}_x}$.
\begin{proof} 
Making some manipulations to (\ref{Kcons}), we have 
\beqn \label{Kcons1}
\begin{array}{l}
 {\mathcal{K'}_\tau } = \left( {1 - \frac{{\left( {1 - x} \right)M}}{x}} \right)\sum\limits_{{n_0} = 1}^N {\left( {\begin{array}{*{20}{c}}
   N  \\
   {{n_0}}  \\
\end{array}} \right){{\left( {1 - {x^M}} \right)}^{{n_0}}}{x^{M\left( {N - {n_0}} \right)}}}  +  \\ 
 \,\,\,\,\,\,\,\,\,\,\,\,\frac{{M\left( {1 - x} \right)}}{{x\left( {1 - {x^M}} \right)}}\sum\limits_{{n_0} = 1}^N {\left( {\begin{array}{*{20}{c}}
   N  \\
   {{n_0}}  \\
\end{array}} \right){n_0}{{\left( {1 - {x^M}} \right)}^{{n_0}}}{x^{M\left( {N - {n_0}} \right)}}}.  \\ 
 \end{array}
\eeqn
It can be observed that $\sum\limits_{{n_0} = 1}^N {\left( {\begin{array}{*{20}{c}}
   N  \\
   {{n_0}}  \\
\end{array}} \right){{\left( {1 - {x^M}} \right)}^{{n_0}}}{x^{M\left( {N - {n_0}} \right)}}} $ and $\sum\limits_{{n_0} = 1}^N {\left( {\begin{array}{*{20}{c}}
   N  \\
   {{n_0}}  \\
\end{array}} \right){n_0}{{\left( {1 - {x^M}} \right)}^{{n_0}}}{x^{M\left( {N - {n_0}} \right)}}} $ represent a cumulative distribution
 function (CDF) and the mean of a binomial distribution \cite{Kris06} with parameter $p$, respectively  missing the term corresponding
  to $n_0 =0$ where $p = 1 - x^M$. Note that the CDF and mean of such a distribution are 1 and 
  $Np = N\left(1-x^M\right)$, respectively. Hence, (\ref{Kcons1}) can be rewritten as
\beqn \label{Kcons2}
{K'_\tau } = \left( {1 - \frac{{\left( {1 - x} \right)M}}{x}} \right)\left( {1 - {x^{MN}}} \right) + \frac{{M\left( {1 - x} \right)}}{{x\left( {1 - {x^M}} \right)}}N\left( {1 - {x^M}} \right).
\eeqn
After some manipulations, we have
\beqn \label{Kcons3}
{\mathcal{K'}_\tau } = 1 - {x^{MN}} + MN{x^{MN - 1}}\left( {1 - x} \right) > 0, \: {\forall x}.
\eeqn
Therefore, we have completed the proof.
\end{proof} 

Hence, the first property of Proposition 1 also holds.

To prove the third property, let us consider the following equation $\frac{{\partial \mathcal{\widetilde{NT}}\left( \tau  \right)}}{{\partial \tau }} = 0$. After some manipulations, we have the following equivalent equation
\beqn \label{gh}
g\left(\tau \right) = h' \left( \tau \right),
\eeqn
where
\beqn \label{gtau}
g\left( \tau  \right) = {\left( {\alpha  + \gamma \sqrt {{f_s}\tau } } \right)^2},
\eeqn 
\beqn \label{hx}
h'\left( \tau  \right) = 2\log \left( {\mathcal{P}\left( {{\mathcal{H}_0}} \right)\gamma \sqrt {\frac{{{f_s}}}{{8\pi }}} \frac{{T - \tau }}{{\sqrt \tau  }}} \right) + {h'_1}\left( x \right),
\eeqn 
\beqn \label{h1x}
{h'_1}\left( x \right) = 2\log \frac{{{K'_\tau }}}{{\sum\limits_{{n_0} = 1}^N {\left( {\begin{array}{*{20}{c}}
   N  \\
   {{n_0}}  \\
\end{array}} \right)f'\left( x \right)} }},
\eeqn
$K'_{\tau}$ is given in (\ref{Kcons}). We have the following result for $h'\left( \tau  \right)$.

\vspace{0.2cm}
\textbf{Lemma 5:} $h'\left( \tau  \right)$ monotonically decreases in $\tau$.

\begin{proof}
The derivative of  $h'\left( \tau  \right)$ can be written as
\beqn \label{heq} 
\frac{{\partial h'}}{{\partial \tau }} = \frac{{ - 1}}{\tau } - \frac{2}{{T - \tau }} + \frac{{\partial {h'_1}}}{{\partial \tau }}.
\eeqn

In the following, we will show that $\frac{{\partial {h'_1}}}{{\partial x}} > 0$ for all $x \in {\mathbb{R}_x}$, all $M$ and $N$, and $\frac{{\partial x}}{{\partial \tau }} < 0$. Hence $\,\frac{{\partial {h'_1}}}{{\partial \tau }} = \frac{{\partial {h'_1}}}{{\partial x}}\frac{{\partial x}}{{\partial \tau }} < 0$. From this, we have $\frac{{\partial h'}}{{\partial \tau }}<0$; therefore, the property stated in lemma 5 holds.

We now show that $\frac{{\partial {h'_1}}}{{\partial x}} > 0$ for all $x \in {\mathbb{R}_x}$, all $M$ and $N$.
Substitute $\mathcal{K'}_\tau$ in (\ref{Kcons3}) to (\ref{h1x}) and exploit the property of the CDF of the binomial distribution function, we have
\beqn \label{h1x1}
\begin{array}{l}
 {h'_1}\left( x \right) = 2\log \frac{{1 - {x^{MN}} + MN{x^{MN - 1}}\left( {1 - x} \right)}}{{\left( {1 - x} \right)\sum\limits_{{n_0} = 1}^N {\left( {\begin{array}{*{20}{c}}
   N  \\
   {{n_0}}  \\
\end{array}} \right){{\left( {1 - {x^M}} \right)}^{{n_0}}}{x^{M\left( {N - {n_0}} \right)}}} }} \\ 
 \,\,\,\,\,\,\,\,\,\,\,\,\, = 2\log \frac{{1 - {x^{MN}} + MN{x^{MN - 1}}\left( {1 - x} \right)}}{{\left( {1 - x} \right)\left( {1 - {x^{MN}}} \right)}} \\ 
 \end{array}.
\eeqn

Taking the first derivative of  ${h'_1}\left( x \right)$ and performing some manipulations, we obtain
\beqn \label{h1x2} 
\frac{{\partial {h'_1}}}{{\partial x}} = 2\frac{{\left( \begin{array}{l}
 r\left( {r - 1} \right){x^{r - 2}}{\left( {1 - x} \right)^2}\left( {1 - {x^r}} \right) \\ 
  + {\left( {1 - {x^r}} \right)^2} + {r^2}{x^{2\left( {r - 1} \right)}}{\left( {1 - x} \right)^2} \\ 
 \end{array} \right)}}{{\left( {1 - {x^r} + r{x^{\left( {r - 1} \right)}}\left( {1 - x} \right)} \right)\left( {1 - x} \right)\left( {1 - {x^r}} \right)}},
\eeqn
where $r = MN$.
It can be observed that there is no negative term in (\ref{h1x2}); hence, $\frac{{\partial {h'_1}}}{{\partial x}} > 0$ for all $x \in {\mathbb{R}_x}$, all $M$ and $N$. Therefore, we have proved the lemma.
\end{proof}

To prove the third property, we show that 
$g\left( \tau \right)$ and $h'\left( \tau \right)$ intersect only once in the range of $\left[ 0, T\right]$. 
This will be done using the same approach as that in Appendix A. Specifically, we will consider two regions
${\mathbf{\Omega}_1}$ and  ${\mathbf{\Omega}_2}$ and prove two properties stated in Lemma 2 for this case.
As in Appendix A, the third property holds if we can prove
$ - \frac{1}{\tau }+\frac{{\partial {h'_1}}}{{\partial \tau }} < \frac{{\partial g}}{{\partial \tau }}$. 
It can be observed that all steps used to prove this inequality are the same as those in the proof of (\ref{partialh1g1ch}) 
for Proposition 1. Hence, we need to prove
\beqn \label{h2lower}
\frac{{\partial {h'_1}}}{{\partial x}} > \frac{2}{{1 - x}}.
\eeqn
Substitute $\frac{{\partial {h'_1}}}{{\partial x}}$ from (\ref{h1x2}) to (\ref{h2lower}), this inequality reduces to
\beqn \label{h2lower1}
2\frac{{\left( \begin{array}{l}
 r\left( {r - 1} \right){x^{r - 2}}{\left( {1 - x} \right)^2}\left( {1 - {x^r}} \right) \\ 
  + {\left( {1 - {x^r}} \right)^2} + {r^2}{x^{2\left( {r - 1} \right)}}{\left( {1 - x} \right)^2} \\ 
 \end{array} \right)}}{{\left( {1 - {x^r} + r{x^{\left( {r - 1} \right)}}\left( {1 - x} \right)} \right)\left( {1 - x} \right)\left( {1 - {x^r}} \right)}}> \frac{2}{1-x}.
\eeqn
After some manipulations, this inequality becomes equivalent to 
\beqn \label{h2lower2}
rx^{\left(r-2\right)}\left(1-x\right)^2\left[r-\left(1+x+x^2+\cdots+x^{\left(r-1\right)}\right)\right]>0.
\eeqn
It can be observed that $0<x<1$ and  $0<x^i<1$, $i \in \left[0,r-1\right]$. Hence, we have $r-\left(1+x+x^2+\cdots+x^{\left(r-1\right)}\right) >0$ which shows that (\ref{h2lower2}) indeed holds. Therefore, (\ref{h2lower}) holds and we have completed the proof of the third property. Finally,
the last property of the Proposition is obviously correct. Hence, we have completed the proof of Proposition 2.

\bibliographystyle{IEEEtran}

%\bibliography{am_ger_eng,rubi_eng}

\newpage

 %Fig. 1

% Fig. 2

% Fig. 4

% Fig. 6

%\begin{figure*}
\begin{table}
\centering
\caption{Comparison between the normalized throughputs of basic access and RTS/CTS access}
\label{table}
\begin{tabular}{|c|c|c|c|c|c|}
\hline 
\multicolumn{6}{|c|}{BASIC ACCESS ($N$,$M$,$m$) = (10,5,4)}\tabularnewline
\hline
\hline 
\multicolumn{2}{|c|}{} & \multicolumn{4}{c|}{$\tau(ms)$}\tabularnewline
\cline{3-6} 
\multicolumn{2}{|c|}{$\mathcal{NT}$} & 1 & 2.6 & 10 & 20\tabularnewline
\hline 
 & 16 & 0.4865 & 0.5545 & 0.5123 &  0.4677\tabularnewline
\cline{2-6} 
 & 64 & 0.5803  & 0.6488  & 0.6004 & 0.5366\tabularnewline
\cline{2-6} 
$W$ & 182 & 0.6053  & \textbf{\emph{\Large 0.6822}} & 0.6323 & 0.5594\tabularnewline
\cline{2-6} 
 & 512 & 0.6014  & 0.6736  & 0.6251 & 0.5567\tabularnewline
\cline{2-6} 
 & 1024 & 0.5744  & 0.6449  & 0.5982 & 0.5312\tabularnewline
\hline
\end{tabular}\begin{tabular}{|c|c|c|c|c|c|}
\hline 
\multicolumn{6}{|c|}{RTS/CTS ACCESS ($N$,$M$,$m$) = (10,5,4)}\tabularnewline
\hline
\hline 
\multicolumn{2}{|c|}{} & \multicolumn{4}{c|}{$\tau(ms)$}\tabularnewline
\cline{3-6} 
\multicolumn{2}{|c|}{$\mathcal{NT}$} & 1 & 2.6 & 10 & 20\tabularnewline
\hline 
 & 16 & 0.6029 & 0.6654 & 0.6236 & 0.5568\tabularnewline
\cline{2-6} 
 & 60 & 0.6022 & \textbf{\emph{\Large 0.6733}} & 0.6231 & 0.5568\tabularnewline
\cline{2-6} 
$W$ & 128 & 0.5954 & 0.6707 & 0.6175 & 0.5533\tabularnewline
\cline{2-6} 
 & 512 & 0.5737  & 0.6444 & 0.5982 & 0.5323\tabularnewline
\cline{2-6} 
 & 1024 & 0.5468 & 0.6134 & 0.5692 & 0.5059\tabularnewline
\hline
\end{tabular}

\begin{tabular}{|c|c|c|c|c|c|}
\hline 
\multicolumn{6}{|c|}{BASIC ACCESS ($N$,$M$,$m$) = (5,3,4)}\tabularnewline
\hline
\hline 
\multicolumn{2}{|c|}{} & \multicolumn{4}{c|}{$\tau(ms)$}\tabularnewline
\cline{3-6} 
\multicolumn{2}{|c|}{$\mathcal{NT}$} & 1 & 2.3 & 10 & 20\tabularnewline
\hline 
 & 16 & 0.5442  & 0.6172  & 0.5647 & 0.5079 \tabularnewline
\cline{2-6} 
 & 64 & 0.6015  & 0.6757  & 0.6302 & 0.5565\tabularnewline
\cline{2-6} 
$W$ & 100 & 0.6094  & \textbf{\emph{\Large 0.6841}}  & 0.6345 & 0.5665\tabularnewline
\cline{2-6} 
 & 512 & 0.5735  & 0.6443  & 0.5983 & 0.5324\tabularnewline
\cline{2-6} 
 & 1024 & 0.5210  & 0.5866  & 0.5447 & 0.4842\tabularnewline
\hline
\end{tabular}\begin{tabular}{|c|c|c|c|c|c|}
\hline 
\multicolumn{6}{|c|}{RTS/CTS ACCESS ($N$,$M$,$m$) = (5,3,4)}\tabularnewline
\hline
\hline 
\multicolumn{2}{|c|}{} & \multicolumn{4}{c|}{$\tau(ms)$}\tabularnewline
\cline{3-6} 
\multicolumn{2}{|c|}{$\mathcal{NT}$} & 1 & 2.5 & 10 & 20\tabularnewline
\hline 
 & 22 & 0.5972 & \textbf{\emph{\Large 0.6789}} & 0.6177 & 0.5529\tabularnewline
\cline{2-6} 
 & 64 & 0.5931 & 0.6674 & 0.6217 & 0.5483\tabularnewline
\cline{2-6} 
$W$ & 128 & 0.5876 & 0.6604 & 0.6131 & 0.5441 \tabularnewline
\cline{2-6} 
 & 512 & 0.5458 & 0.6128 & 0.5691 & 0.5057\tabularnewline
\cline{2-6} 
 & 1024 & 0.4965 & 0.5591 & 0.5189 & 0.4610\tabularnewline
\hline
\end{tabular}
\end{table} 
%\end{figure*}

%thebibliography
% biography section
% 
% If you have an EPS/PDF photo (graphicx package needed) extra braces are
% needed around the contents of the optional argument to biography to prevent
% the LaTeX parser from getting confused when it sees the complicated
% \includegraphics command within an optional argument. (You could create
% your own custom macro containing the \includegraphics command to make things
% simpler here.)
%\begin{biography}[{\includegraphics[width=1in,height=1.25in,clip,keepaspectratio]{mshell}}]{Michael Shell}
% or if you just want to reserve a space for a photo:

%\begin{IEEEbiography}{Michael Shell}
%Biography text here.
%\end{IEEEbiography}
%
%% if you will not have a photo at all:
%\begin{IEEEbiographynophoto}{John Doe}
%Biography text here.
%\end{IEEEbiographynophoto}
%
%% insert where needed to balance the two columns on the last page with
%% biographies
%%\newpage
%
%\begin{IEEEbiographynophoto}{Jane Doe}
%Biography text here.
%\end{IEEEbiographynophoto}

% You can push biographies down or up by placing
% a \vfill before or after them. The appropriate
% use of \vfill depends on what kind of text is
% on the last page and whether or not the columns
% are being equalized.

% that's all folks
\end{document}